\documentclass[twocolumn,aps,prd,   
               preprintnumbers,numbers,sort&compress,
               nofootinbib,
                            showpacs,
               colorlinks,
               linkcolor=blue,
               citecolor=blue]{revtex4-1} 

\usepackage{graphicx}
\usepackage{dcolumn}
\usepackage{bm}
\usepackage{hyperref}

\usepackage{amsmath}
\usepackage[caption=false]{subfig}

 \def\ra{\rangle}
\def\la{\langle}

\newcommand{\be}{\begin{eqnarray}}
\newcommand{\ee}{\end{eqnarray}}
\newcommand{\beq}{\begin{equation}}
\newcommand{\eeq}{\end{equation}}
\newcommand{\exclude}[1]{}

\newcommand{\jcap}{JCAP}

\newcommand{\aap}{Astronomy \& Astrophysics}
\newcommand{\mnras}{Mon.\ Not.\ R.\ Astron.\ Soc.}
\newcommand{\apjs}{Astrophys.\ J.\ Supp.}
\newcommand{\solphys}{Sol.\ Phys.}

\graphicspath{{./Figures/}} 

\begin{document}
       \title{Impulsive radio events in quiet solar corona and Axion Quark Nugget Dark Matter}
       \author{Shuailiang Ge}
       \email{slge@phas.ubc.ca}
        \affiliation{Department of Physics and Astronomy, University of British Columbia, Vancouver, V6T 1Z1, BC, Canada}
        \author{Md Shahriar Rahim Siddiqui}
        \email{shahriar.naf07@gmail.com}
       \author{Ludovic Van Waerbeke}
       \email{waerbeke@phas.ubc.ca}
        \affiliation{Department of Physics and Astronomy, University of British Columbia, Vancouver, V6T 1Z1, BC, Canada}
       \author{Ariel  Zhitnitsky}
       \email{arz@phas.ubc.ca}
       \affiliation{Department of Physics and Astronomy, University of British Columbia, Vancouver, V6T 1Z1, BC, Canada}
     
       \begin{abstract}
    The Murchison Widefield Array (MWA) recorded \cite{Mondal-2020} impulsive radio events in the quiet solar corona at frequencies 98, 120, 132, and 160 MHz.
    We propose that these radio events are the direct manifestation of dark matter annihilation events within  the axion quark nugget (AQN) 
    framework. It has been argued  \cite{Zhitnitsky:2017rop,Raza:2018gpb} that the AQN annihilation events in the quiet solar corona can be identified with the nanoflares conjectured  by Parker \cite{Parker-1983}. We further support this claim by demonstrating that observed impulsive radio events  \cite{Mondal-2020}, including their rate of appearance, their temporal and spatial distributions and their energetics,  are matching the generic consequences of AQN annihilations in the quiet corona. We propose to test this idea by analyzing the correlated clustering of impulsive radio events in different frequency bands.  These correlations are expressed in terms of the time delays between radio events in different frequency bands, measured in seconds. We also make generic predictions for low (80 and  89 MHz)  and high (179, 196, 217 and 240 MHz) frequency bands, that have been recorded, but not published, by  \cite{Mondal-2020}. We finally suggest to test our proposal by studying possible cross-correlation between MWA radio signals and Solar Orbiter recording of extreme UV photons (a.k.a. ``campfires").
 
       \end{abstract}
     
       \maketitle

\section{Introduction}\label{sec:introduction}
In this work, we discuss two seemingly unrelated stories. The first one is motivated by the recent observations of impulsive radio events in the quiet solar corona (at 98, 120, 132, and 160 MHz), carried out by the Murchison Widefield Array (MWA) \cite{Mondal-2020}.
 The second one is the axion quark nugget (AQN) dark matter model \cite{Zhitnitsky:2002qa} and its possible role in the heating of the solar corona \cite{Zhitnitsky:2017rop,Raza:2018gpb}. The topic of the present paper is to explain how, and why, these two stories are related. 
 
The observed impulsive radio events \cite{Mondal-2020} appear to have all the features normally attributed to nanoflares, conjectured by Parker \cite{Parker-1983}) as a possible resolution of the corona heating mystery \cite{Grotrian-1939}. On the other hand, \cite{Zhitnitsky:2017rop,Raza:2018gpb} have shown that AQNs entering the Sun's corona lead to impulsive energy injection events, that provide the proper amount of energy needed to heat the corona. This led to the identification of AQN annihilation events with nanoflares. Furthermore, most of the AQN-annihilation energy was shown to be released in the transition region, at an altitude around 2000 km, a region known to be the most puzzling layer of the solar corona, where the temperature and the density of the plasma experience a dramatic change across a thin layer. We will show that the annihilation events proposed in \cite{Zhitnitsky:2017rop,Raza:2018gpb} share many features with the impulsive radio signals observed by \cite{Mondal-2020}, in terms of their rate of appearance, their temporal and spatial distributions, their energetics, and other related observables.

Let us first start with an overview of the solar corona heating puzzle. The solar photosphere is in thermal equilibrium at $\sim 5800$ K, while the corona has a temperature of a few $10^6$ K \cite{Grotrian-1939}. Physically, this high temperature corresponds to an energy excess of a few $10^{27}~{\rm erg~s^{-1}}$, mostly observed  in the extreme ultraviolet (EUV) and soft X-ray bands. The conventional view is that the corona excess heating is explained by nanoflares, a concept originally invented by Parker \cite{Parker-1983}. The individual short energy bursts associated with these nanoflares are significantly below detection limits and have not yet been observed in the EUV or  X-ray regimes. In fact, all coronal heating models advocated so far seem to require the existence of an unobserved (i.e. unresolved with current instrumentation) source of energy distributed over the entire Sun \cite{De-Moortel-2015}. Therefore, nanoflares are modelled as invisible generic events, producing an impulsive energy release at a small scale; their cause and their nature are not specified (\cite{Klimchuk:2005nx,Klimchuk:2017}). \cite{Mondal-2020} adopted this definition of nanoflares to explain the impulsive radio events they observed in the quiet solar corona (in terms of frequency of appearance, duration, and wait times distribution, at frequencies 98, 120, 132, and 160 MHz). They argued that radio observations allow to probe much weaker energy levels, with much better temporal and spatial resolutions, in comparison to the current generation instrumentation in EUV and X-rays energy bands. In other words, radio observations can potentially "see" individual nanoflares and their "internal structures", where more energetic EUV and X-rays instruments cannot. 

Second, let us highlight the basic features of the AQN model, while deferring a more detailed overview to Section \ref{AQN-flux}.
The  axion quark nugget (AQN) dark matter  model  \cite{Zhitnitsky:2002qa} was  invented long ago with the single objective of explaining the proximity of the dark and the visible matter densities in the Universe, i.e. $\Omega_{\rm DM}\sim \Omega_{\rm visible}$, without fine tuning. The AQN model construction is, in many respects, similar to the original quark-nugget model suggested by Witten \cite{Witten:1984rs} (see  \citep{Madsen:1998uh} for a review). This type of DM  is "cosmologically dark'', not because of the weakness of AQN interactions, but because of their small cross-section-to-mass ratio, which scales down many observables. 

Two additional elements of the AQN model make it a viable DM model compared to the original proposal \cite{Witten:1984rs,Madsen:1998uh}. First, axion domain walls provide a stabilization factor to AQNs. They are copiously produced during the QCD phase transition, which helps alleviating a number of stability problems with the original nugget model. Secondly, axion quark nuggets can be made of {\it matter} as well as {\it antimatter} during the QCD transition. Consequently, the DM density $\Omega_{\rm DM}$ and the baryonic matter density $\Omega_{\rm visible}$ automatically assume the same order of magnitude ($\Omega_{\rm DM}\sim \Omega_{\rm visible}$), without any fine tuning \cite{Zhitnitsky:2002qa}. One should emphasize that AQNs are stable over cosmological time scales. Antimatter, hidden in the form of these very dense anti-nuggets, is unavailable for annihilation, unless an anti-nugget hits a star or a planet. Very rare annihilation events also happen in the center of galaxies, via collisions with single protons, electrons or light nuclei. They may explain some of the observed excess emission in our Galaxy, in different frequency bands (see next Sect. \ref{AQN-flux} for  references). We will only focus on AQNs made of antimatter, the ones capable of releasing a significant amount of energy via annihilation when they enter the solar corona. As noted by \cite{Zhitnitsky:2017rop}, the power required to solve the corona EUV excess is of the order of $10^{27}~ {\rm erg~s^{-1}}$. This corresponds to the $mc^2$ energy available from DM falling on the Sun (by gravity only), assuming  a typical DM mass density of $\rho_{DM} \simeq 0.3 \ {\rm GeV\,cm^{-3}}$. This correspondence motivated the identification of nanoflares with AQN annihilation events. Furthermore, \cite{Raza:2018gpb} showed that the dominant portion of the annihilation energy is deposited in the corona, before entering the dense regions of the photosphere, at an altitude of approximately 2000 km, known as the Transition Region.

The main goal of the present work is to explore the possibility that AQN's annihilations and nanoflares are the same impulsive energetic events, and the interpretation of the impulsive radio events observed by \cite{Mondal-2020}. Our presentation is organized as follows. In section \ref{AQN-flux}, we overview the basic ideas of the AQN model in the context of the impulsive radio events. In section \ref{corona}, we highlight some features related to the solar corona heating, within the AQN framework.  In sections \ref{radio} and \ref{wait_times}, we present our estimates supporting the main claim of this work, i.e. that observations  \cite{Mondal-2020}  nicely match the characteristics of AQN annihilation events, including the frequency of appearance, the temporal and spatial distributions,  the energetics,  and other related  observables in radio frequency bands.  
 
\section{The AQN model: the basics}\label{AQN-flux}

	It is commonly  assumed that the Universe began in a symmetric state with zero global baryonic charge, and later (through some baryon number violating process, non- equilibrium dynamics, and $\cal{CP}$ violation effects, realizing three famous Sakharov's criteria) evolved into a state with a net positive baryon number. This is called "baryogenesis''. 

	The original motivation for the AQN model comes from the possibility of an alternative to this scenario, where "baryogenesis'' is replaced by a charge separation process in which the global baryon number of the universe remains zero at all times.  In this model, the unobserved anti-baryons come to comprise dark matter in the form of dense anti-nuggets, made of antiquarks and antigluons, in a colour superconducting (CS) phase. This "charge separation" process results in two populations of AQNs, carrying positive or negative baryon numbers. In other words,  an AQN can be formed of either {\it matter or antimatter}. However, due to the global  $\cal CP$ violating processes associated with $\theta_0\neq 0$, during the early formation stage, the number of nuggets and anti-nuggets was different\footnote{. This source of strong ${\cal CP}$ violation is no longer available at the present epoch, as a result of the dynamics of the axion, which  remains the most compelling resolution of the strong ${\cal CP}$ problem (see \cite{axion1,axion2,axion3,KSVZ1,KSVZ2,DFSZ1,DFSZ2} and recent reviews \cite{vanBibber:2006rb,Asztalos:2006kz,Sikivie:2008,Raffelt:2006cw,Sikivie:2009fv,Rosenberg:2015kxa,Marsh:2015xka,Graham:2015ouw, Irastorza:2018dyq}.}.) This difference is always an order one effect, irrespective of the parameters of the theory (i.e. the axion mass $m_a$, or the initial misalignment angle $\theta_0$). We refer to the original papers \cite{Liang:2016tqc,Ge:2017ttc,Ge:2017idw,Ge:2019voa} devoted to axion quark nuggets' formation, generation of the baryon asymmetry, and survival pattern of nuggets during the evolution in  early Universe.
    
	Antimatter AQNs can interact with regular matter via annihilations, which leads to electromagnetic radiations whose spectral characteristics and flux can be calculated within the AQN framework. These emissions are sufficiently dim to not violate any known observational constraints, but are strong enough to offer a possible solution to some unexplained astrophysical observations. For instance, it is known that the galactic spectrum contains several excesses of diffuse emission, the best known example being the strong galactic 511~keV line, the origin of which is not well established and remains debated. If AQNs have a baryon number in the $\langle B\rangle \sim 10^{25}$ range, they can offer a potential explanation for several of these diffuse components, in three different spectral domains : radio, X-ray and $\gamma$-ray. In all three cases, the photon emission originates from the outer layer of the AQN, known as the electrosphere. All intensities in different frequency bands are expressed in terms of a single parameter, $\langle B\rangle$, such that all the relative intensities are unambiguously fixed, because they are determined by the Standard Model (SM) of particle physics. This constitutes a nontrivial consistency check of the AQN model. For further details, see the original works \cite{Oaknin:2004mn, Zhitnitsky:2006tu,Forbes:2006ba, Lawson:2007kp,Forbes:2008uf,Forbes:2009wg}.
    
	Interestingly, AQNs could also offer a resolution to other seemingly unrelated puzzles, such as the "Primordial Lithium Puzzle" \cite{Flambaum:2018ohm} or the annual modulation observed by DAMA/LIBRA  (see \cite{Zhitnitsky:2019tbh}). Furthermore, it may resolve the puzzling seasonal variation of the X-ray background in the near-Earth environment, in the 2-6 keV energy range \citep{Fraser:2014wja}, as suggested in \cite{Ge:2020cho}. AQN annihilation events could also explain a mysterious type of explosions in the Earth's atmosphere, where infrasonic and seismic acoustic waves have been recorded, without any traces of accompanying meteor-like events (\cite{Budker:2020mqk}). Finally, AQN annihilation events which occur under a thunderstorm may explain several events observed by the Telescope Array collaboration, as discussed in \cite{Zhitnitsky:2020shd}. In the context of our study, however, the most important application of AQNs is a possible explanation of the  solar corona heating (  \cite{Zhitnitsky:2017rop,Raza:2018gpb}), which is reviewed in details in next subsection \ref{corona}. 
    
	The key parameter, which determines the intensity of the effects mentioned above, is the average baryon charge $\la B \ra$ of the AQNs. It is expected that AQNs do not have all the same $B$, but rather $B$ is given by a distribution function, $f(B)$. There are several constraints on this parameter which are reviewed below. AQNs are macroscopically large objects, with a typical size of $R\simeq 10^{-5}{\rm cm}$. They have roughly nuclear density, resulting in masses of roughly 10 grams. For an AQN with a baryonic charge $B$, mass is given by $M_N\simeq m_p |B|$. For the present work, we adopt a typical nuclear density of order $10^{40}\,{\rm g ~cm^{-3}}$, such that a nugget with $|B|\simeq 10^{25}$ has a typical radius $R\simeq 10^{-5}{\rm cm}$. One can view an AQN as a very small neutron star (NS), with nuclear density. The difference is that a NS is squeezed by gravity, while an AQN is squeezed by axion domain wall pressure. 
    
	Let us now overview the observational constraints on AQNs. The strongest direct detection limit \footnote{There is also an indirect constraint on the flux of  dark matter nuggets with mass $M< 55$g (which corresponds approximately $B\simeq 10^{25}$) based on the non-detection of etching tracks in ancient mica \cite{Jacobs:2014yca}. It slightly touches the lower bound of the allowed  range (\ref{direct}), but does not strongly constraint the entire window (\ref{window}) because the dominant portion of AQNs lies well above its lower limit (\ref{direct}),  assuming the mass distribution \ref{f(B)})} is set by the Ice Cube Observatory's observations (see Appendix A in \cite{Lawson:2019cvy}):
\be
\label{direct}
\la B \ra > 3\cdot 10^{24} ~~~[{\rm direct ~ (non)detection ~constraint]}.
\ee

The authors of \cite{Herrin:2005kb} use the Apollo data to constrain the abundance of AQNs, in the region of 10\,kg to one tonne. It has been argued that the contribution of such heavy nuggets  must be at least an order of magnitude less than would saturate the dark matter in the solar neighbourhood \cite{Herrin:2005kb}. Assuming that AQNs do saturate dark matter, the constraint  \cite{Herrin:2005kb} can be reinterpreted as at least $90\%$ of the AQNs having masses below 10\,kg. This constraint can be approximately expressed  in terms of the baryon charge:
   \be
\label{apollo}
\la B \ra \lesssim   10^{28} ~~~ [  {\rm   Apollo~ constraint ~} ].
\ee
Therefore, indirect observational constraints (\ref{direct}) and (\ref{apollo}) suggest that, if the AQNs exist, and saturate the dark matter density today, the dominant portion of them must reside in the window: 
\be
\label{window}
3\cdot 10^{24}\lesssim\la B \ra \lesssim   10^{28}~ [{\rm constraints~ from~ observations}].  ~~~
\ee 
 
The authors of \cite{SinghSidhu:2020cxw} considered a generic constraints for nuggets made of antimatter (ignoring all essential specifics of the AQN model, such as quark matter phase of the nugget's core). Our constraints (\ref{window}) are consistent with their findings, including the Cosmic Microwave Background (CMB) and Big Bang Nucleosynthesis (BBN) and others, with the exception of "Human Detectors" \footnote{We think that the corresponding estimates of \cite{SinghSidhu:2020cxw} are  oversimplified, and do not have the same status as those derived from CMB or BBN constraints. In particular, the rate of energy deposition was estimated in \cite{SinghSidhu:2020cxw} assuming that the annihilation processes between anti-nuggets and baryons are similar to $p\bar{p}$ annihilation process. It is known that it cannot be the case in general, and it is not the case in particular in the AQN model because the annihilating objects have drastically different structures. It has been also assumed in \cite{SinghSidhu:2020cxw} that  a typical X-ray energy  is around  1 keV, which is  much lower than direct computations in the AQN model would  suggest \cite{Budker:2020mqk}. Higher energy x-rays have much longer mean-free path, which implies that the dominant portion of the energy will be deposited outside the human body. Finally, \cite{SinghSidhu:2020cxw} assume that an anti-nugget will result in an "injury similar to a gunshot". It is obviously a wrong picture as the size of a typical nugget is only $R\sim 10^{-5} {\rm cm}$ while the most of the energy is deposited in form of the x-rays on centimeter  scales \cite{Budker:2020mqk} without making a large hole similar to  bullet as assumed in \cite{SinghSidhu:2020cxw}. In this case a human's  death may occur as a result of  a large dose of radiation with a long time delay, which would make  it hard to identify the cause of the death.}.
 
We emphasize that the AQN model, within the above window (\ref{window}), is consistent with all presently available cosmological, astrophysical, satellite and ground-based constraints.  Furthermore, it has been shown that these macroscopic objects can be formed, and the dominant portion of them will survive highly disruptive events (such as BBN, galaxy and star formation) during the long evolution of the Universe \cite{Liang:2016tqc,Ge:2017ttc,Ge:2017idw,Ge:2019voa}. The AQN model is very rigid, and predictive, as there is no flexibility, nor freedom to modify any estimates \cite{Oaknin:2004mn, Zhitnitsky:2006tu,Forbes:2006ba, Lawson:2007kp,Forbes:2008uf,Forbes:2009wg,Flambaum:2018ohm,Zhitnitsky:2017rop,Raza:2018gpb,Zhitnitsky:2019tbh,Ge:2020cho,Budker:2020mqk,Zhitnitsky:2020shd}, which have been carried out in drastically different environments, where  densities and  temperatures span  many orders in magnitude.  

\section{The AQN model:  application to the solar corona heating}\label{corona} 
In this section, we overview the basic characteristics of nanoflares, from an AQN viewpoint. The corresponding results will play a vital role in our studies  in  section \ref{radio},  where we interpret the   radio events analyzed by \cite{Mondal-2020}  in terms of  the AQN annihilation events \cite{Zhitnitsky:2017rop,Raza:2018gpb}.   
\subsection{The nanoflares: observations and modelling}\label{nanoflares} 
We start with a few historical remarks. The solar corona is a very peculiar environment. Starting at an altitude of 1000 km above of the photosphere, the highly ionized iron lines show that the plasma temperature exceeds a few $10^6$ K. The total energy radiated away by the corona is of the order of $L_{\rm corona} \sim 10^{27} {\rm erg~ s^{-1}}$, which is about $10^{-6}-10^{-7}$ of the total energy radiated by the photosphere. Most of this energy is radiated at the extreme ultraviolet (EUV) and soft X-ray wavelengths. There is a very sharp transition region, located in the upper chromosphere, where the temperature suddenly jumps from a few thousand degrees to $10^6$ K. This transition layer is relatively thin, 200 km at most. This transition happens uniformly over the Sun, even in the quiet Sun, where the magnetic field is small ($\sim 1~ {\rm G}$), away from active spots and coronal holes. The reason for this uniform heating of the corona is not understood.

A possible solution to the heating problem in the quiet Sun corona was proposed in 1983 by Parker \citep{Parker}, who postulated that a continuous and uniform sequence of miniature flares, which he called ``nanoflares'', could happen in the corona. This became the conventional view. The term "nanoflare" has been used in a series of papers by Benz and coauthors
  \cite{Benz-2000,Benz-2001, Kraev-2001,Benz-2002, Benz-2003}, and many others, to advocate the idea that these small "micro-events'' might be  responsible for the   heating of the   quiet solar corona. We   want to  mention a few relatively  recent studies   \cite{Pauluhn:2006ut,Hannah:2007kw,Bingert:2012se,terzo-2011,Bradshaw-2012,Jess-2014,Kirichenko-2017,Cormack-2017} and reviews \cite{Klimchuk:2005nx,Klimchuk:2017} which support the basic  claim of earlier works, i.e. that nanoflares play the  dominant  role in the heating of the solar corona.

 In what follows, we adapt  the definition suggested in \cite{Benz-2003} and refer to nanoflares as ``micro-events" in quiet regions of the corona, to be contrasted with ``micro flares," which are significantly larger in scale and observed in active   regions.   The term ``micro-events" refers to a short enhancement of coronal emission in the energy range of about $(10^{24}-10^{28})$erg. One should emphasize that the lower limit   gives the instrumental threshold for observing quiet  regions, while the upper limit refers to the smallest events observable in active regions. The list below shows the most important constraints on 
nanoflares from the observations of the EUV iron lines with SoHO/EIT:\\
 {\bf 1.} The EUV emission is highly isotropic \citep{Benz-2001, Benz-2002}, therefore the nanoflares have to be distributed very ``uniformly in quiet regions'', in contrast with micro-flares and flares
 which are much more energetic and occur exclusively in active areas \citep{Benz-2003}. For instance, flares have a highly non-isotropic spatial distribution because they are associated with the active regions;\\
{\bf 2.} According to \cite{Kraev-2001}, in order to reproduce the measured EUV excess, the observed range of nanoflares  needs to be extrapolated from the observed events  interpolating between $(3.1\cdot 10^{24}  - 1.3\cdot 10^{26})~{\rm erg}$ to sub-resolution events with much smaller energies, see item 3 below.\\
 {\bf 3.} In order to reproduce the measured  radiation loss, the observed range of nano flares (having a lower limit at about $3\cdot 10^{24}$erg due to the instrumental threshold) needs to be extrapolated to  energies as low as  $ 10^{22}$erg  and in some models, even to $ 10^{20}$erg (see table 1 in \cite{Kraev-2001});\\
{\bf 4.} The nanoflares and microflares appear in a different range of temperature and emission measure (see  Fig.3 in \cite{Benz-2003}). While  the instrumental  limits prohibit observations at intermediate temperatures, nevertheless the authors of \cite{Benz-2003} argue that  ``the occurrence rates of nanoflares and microflares are so different that they cannot originate from the same population". We emphasize this difference to argue that the flares originate at sunspot areas, with locally large magnetic fields ${\cal{B}}\sim (10^2-10^3)$ G, while  the EUV emission (which is observed  even in very quiet regions where ${\cal{B}}\sim 1$G) is isotropic and covers the entire solar surface;\\
{\bf 5.} Time measurements of many nanoflares demonstrate a Doppler shift with typical velocities of (250-310) km/s (see Fig.5 in \cite{Benz-2000}).
The observed line width in OV  of $\pm 140$ km/s far exceeds the thermal ion velocity, which is around 11 km/s   \cite{Benz-2000};\\
 {\bf 6.} The temporal evolution of flares and nanoflares also appears different. The typical ratio between the maximum and minimum EUV irradiance during  the solar cycle does not exceed a factor of 3 between its maximum in 2000 and its minimum in 2009 (see Fig. 1 from \cite{Bertolucci-2017}), while the same ratio for flares and sunspots is much larger, of the order of $10^2$.
If the magnetic reconnection was fully responsible for both the flares and nanoflares, then  the variation during the solar cycles should be similar for these two phenomena. It is not what is observed; the modest variation of the EUV with the solar cycles in comparison to the flare fluctuations suggests  that the EUV radiation does not directly follow the magnetic field activity, and that the EUV fluctuation is a  secondary, not a primary effect of the magnetic activity.

 The nanoflares are usually characterised by the following distribution:
 \begin{equation}
\label{dN}
 d N\propto   W^{-\alpha} d W ~~~~ 10^{21}{\rm erg}
 \lesssim W\lesssim 10^{26} {\rm erg}
\end{equation}
 where $ d N$ is the number of nanoflare events per unit time, with an energy between $W$ and $W+ d W$. 
 In formula (\ref{dN}), we display  the approximate energy window for $W$ as expressed by items {\bf 2} and {\bf 3}, including the sub-resolution events extrapolated to very low energies. 
 The distribution   $\rm d N/\rm d W$ has been  modelled via magnetic-hydro-dynamics (MHD) simulations \cite{Pauluhn:2006ut,Bingert:2013} in such a way that the Solar observations match the simulations. The parameter $\alpha$ was  fixed to fit  observations \cite{Pauluhn:2006ut,Bingert:2013}, (see the description of the different models in next subsection).

 \subsection{The nanoflares as AQN annihilation events}\label{identification} 
 It has been conjectured in  \cite{Zhitnitsky:2017rop} that the nanoflares can be identified with AQN annihilation events. This conjecture was essentially motivated by the fact that the amount of energy available from the dark matter falling on the Sun per second, in the form of mass ($mc^2$) , is similar to the amount of energy needed to maintain the corona at its observed temperature ($\sim 10^{27}~ {\rm erg~s^{-1}}$). The dark matter density in the solar system is estimated to be of the order of $\rho_{\rm DM} \simeq 0.3 \ {\rm GeV\,cm^{-3}}$, within a factor $\sim 2$. From this identification, it follows that the  baryon charge distribution (within the AQN framework) and the nanoflare energy distribution (\ref{dN}) must be one and the same function \cite{Zhitnitsky:2017rop}, i.e.
  \begin{equation}
\label{W_B}
 d N\propto B^{-\alpha} d B\propto W^{-\alpha} d W
\end{equation}
where $d N$ is the number of nanoflare events  with energy between $W$ and $W+ d W$, which occur as a result of the complete annihilation of the antimatter AQN carrying a baryon charges between $B$ and $B+ d B$.  

An immediate self-consistency check of this conjecture is the observation that the allowed window (\ref{window}) for the AQNs baryonic charge largely overlaps with  the approximate energy window for nanoflares, $W$ expressed by (\ref{dN}). 
 This is because the  annihilation of a single baryon charge produces an energy of about $2m_pc^2\simeq 2~ {\rm GeV}$, which can be expressed in terms of the conventional units as follows,
 \be
 \label{units}
 1~ {\rm GeV}=1.6 \cdot 10^{-10} {\rm J}=1.6\cdot 10^{-3} {\rm erg}, 
 \ee 
 such that the nanoflare energy $W$ for the anti-nugget with baryon charge $B$ can be approximated as  $W\simeq 2~ {\rm GeV\cdot }B$. One should emphasize that this is a highly nontrivial self-consistency check of proposal  \cite{Zhitnitsky:2017rop}, as the acceptable windows (\ref{window}) and  (\ref{dN}) for the AQNs and nanoflares have been constrained from drastically different  physical systems. 
  
  Encouraged by this self-consistency check and the highly nontrivial energetic consideration, \cite{Raza:2018gpb} used the power-law index $\alpha$ entering (\ref{dN}) to describe the baryon number distribution 
 $dN/dB$ for the anti-nuggets,  which represents the direct consequence\footnote{One should   note that it has been  argued \cite{Ge:2019voa} that the algebraic scaling (\ref{W_B}) is a generic feature of the AQN formation mechanism based on percolation theory. The phenomenological parameter $\alpha$ is determined by the properties of the domain wall formation during the QCD transition in the early Universe, but it cannot be theoretically computed in strongly coupled QCD. Instead, it will be constrained based on the observations as discussed in the text.} of the  conjecture (\ref{W_B}). 
 More specifically, in the Monte Carlo (MC) simulations performed in \cite{Raza:2018gpb} , the baryon number distribution of the AQNs, as given by (\ref{f(B)}) , is assumed to directly follow the nanoflare distribution $dN/dW$ , with the same index $\alpha$ as  the conjecture (\ref{W_B}) states. 
 
 The nanoflare distribution models proposed  in \cite{Pauluhn:2006ut,Bingert:2013} have been  adapted by  \cite{Raza:2018gpb}.  Three different choices for the power-law index $\alpha$ have been considered in  \cite{Pauluhn:2006ut,Bingert:2013}:
\begin{equation}
\label{eq:2.2 f(B) ass_alpha}
\alpha=2.5,~2.0,~{\rm or}\left\{
\begin{aligned}
&1.2  &W\lesssim 10^{24} {\rm erg} \leftrightarrow B\lesssim 3\times 10^{26}  \\
&2.5  &W\gtrsim 10^{24} {\rm erg}  \leftrightarrow B\gtrsim  3\times 10^{26}. \nonumber
\end{aligned}\right.
\end{equation}
In addition to the power law index $\alpha$, different models are also characterized by different choices of $B_{\rm min}$: $10^{23}$ and $3\times10^{24}$. Therefore, a total of 6 different models  have been discussed in \cite{Pauluhn:2006ut,Bingert:2013} which we expressed in terms of the baryon charge $B$ rather than in terms of the nanoflare energy $W$. We also fix $B_{\rm max}=10^{28}$ to be consistent with the constraint (\ref{window}). 

In this work, we  will only use simulations with $\langle B\rangle\gtrsim10^{25}$ in order to be consistent with (\ref{window}). This means that we are excluding two models considered in \cite{Pauluhn:2006ut,Bingert:2013}: the one with $B_{\rm min} \sim 10^{23}$ and the one with $\alpha=2, 5$ and $\alpha=2$. We also exclude the model with $B_{\rm min} \sim 10^{23}$  and that with $\alpha=1.25$ and $\alpha=2.5$ to simplify things as it produces results very similar to another model.   The remaining three models are labeled as follows:
\be
\label{eq:groups}
    &&{\rm Group~1:~} B_{\rm min}=3\times 10^{24}, \alpha=2.5 \\
    &&{\rm Group~2:~} B_{\rm min}=3\times 10^{24}, \alpha=2.0 \nonumber \\
    &&{\rm Group~3:~} B_{\rm min}=3\times 10^{24}, \alpha= \begin{cases}1.2,~B\lesssim 3\times 10^{26}\\ 2.5, ~B\gtrsim 3\times 10^{26} \end{cases}
      \nonumber
\ee
while $B_{\rm max}=10^{28}$ for all the models.

\exclude{
In Table \ref{table:2.2 mean B} we show the mean baryon charge $\langle B \rangle$ for each of the 6 models.
\begin{table} [h] 
	\caption{Values of the mean baryon charge $\langle B\rangle$ for different parameters of the AQN mass-distribution function.} 
	\centering 
	\begin{tabular}{c|ccc} 
		\hline\hline
		$(B_{\rm min},\alpha)$ & 2.5                 & 2.0                 & (1.2, 2.5)          \\ \hline
		$10^{23}$                   & $2.99\times10^{23}$ & $1.15\times10^{24}$ & $ {4.25\times10^{25}}$ \\ 
		$3\times10^{24}$                   & $8.84\times10^{24}$ & $2.43\times10^{25}$ & $ {1.05\times10^{26}}$ \\ \hline
	\end{tabular}
	\label{table:2.2 mean B} 
\end{table}
For the simulations in  \cite{Raza:2018gpb} and in this work, we will only investigate parameters that give $\langle B\rangle\gtrsim10^{25}$  to be consistent with (\ref{window}). This means that we are excluding two models considered in \cite{Pauluhn:2006ut,Bingert:2013}: that with $B_{\rm min} \sim 10^{23}$ and that with $\alpha=2, 5$ and $\alpha=2$.
} 
The average baryon number of the distribution is defined as
\begin{equation}
\label{f(B)}
\langle B\rangle 
=\int_{B_{\rm min}}^{B_{\rm max}}\rm d B~[B ~ f(B)],
~~~~\frac{ dN}{dB}\propto     f(B)\propto B^{-\alpha}
\end{equation}
where $\rm f(B)$ is normalized and the power-law is taken to hold in the range from $B_{\rm min}$ to $B_{\rm max}$.   

The above estimate reveals an astonishing coincidence between the energy/mass windows (\ref{window}) and  (\ref{dN}) for AQNs and nanoflares respectively. This coincidence is a strong support of our proposal \cite{Zhitnitsky:2017rop,Raza:2018gpb} that the nanoflares and the AQN annihilation events are the same phenomena (see items {\bf 2} and {\bf 3} of Section \ref{nanoflares}).

We are now in position to present several additional arguments in favor of our proposal: item {\bf 1} (Section \ref{nanoflares}) is also naturally explained in the AQN framework as DM is expected to be distributed very uniformly over the Sun, making no distinction between quiet and active regions, in contrast with large flares. A similar argument applies to item {\bf 4} ,  as the strength of the magnetic field and its localization is absolutely irrelevant for the nanoflare events in form of the AQNs, in contrast with conventional paradigm where nanoflares are thought to be scaled down configurations of their larger cousins,  which are much more energetic and occur exclusively in active areas and cannot be uniformly distributed. 
 
 The existence of a large Doppler shift,  with a typical velocities (250-310) km/s, mentioned in  item {\bf 5}, can be  understood  within the AQN interpretation as the following: the typical velocities of an anti-nugget entering the solar corona is very high, around 700 km/s.  The Mach number  $M=v_{\rm AQN}/c_s$  is also very large. A shock wave will be formed and will push the surrounding material to velocities which are much higher than would normally be present at thermal  equilibrium. 
 
 Finally, as stated in item {\bf 6}, the temporal modulation of the EUV irradiance over a solar cycle is very small and does not exceed a factor $\sim 3$, as opposed to the much dramatic changes in Solar activity, with modulations on the level of $10^2$ over the same time scale. This suggests that the energy injection from the nanoflares is weakly related to solar activity, which is
in contradiction with the picture where magnetic reconnection modulated by the Sun activity plays an essential role in the formation and dynamics of nanoflares. This is, however, consistent with our interpretation of nanoflares being associated with AQN annihilation events, as an external cause of the main source of the EUV irradiance.

\section{The AQN model   confronts the radio observations }\label{radio} 
We start in subsection \ref{mechanism} by describing the basic mechanism of the radio emission due to AQN annihilation events in the solar corona.
We estimate the event rate in subsection \ref{rate}. The role of non-thermal electrons in generation of the radio signal events is discussed in subsection \ref{non-thermal}. Finally, in subsection \ref{energetics}
we estimate the intensity of the radio signal events. 

\subsection{Mechanism of the radio emission in solar corona.}\label{mechanism} 
It is generally accepted that the radio emission from the corona results from the interaction of plasma  oscillations (also known as Langmuir waves) with non-thermal electrons which must be injected into the plasma \cite{Thejappa-1991}. An important  element for the successful  emission of radio waves is that a plasma instability must develop.   It occurs when the injected electrons have a  non-thermal high energy component,  with a momentum distribution function characterized by a positive derivative\footnote{If the derivative has a negative sign it will lead to the so-called Landau damping.}  with respect to the electron's velocity. In this case, a plasma instability develops and radio waves  can be emitted. 

The frequency of emission $\nu$ is mostly determined by the plasma frequency $\omega_p$ in a given environment, i.e. 
 \begin{equation}
\label{eq:omega}
 \omega^2=\omega_p^2+ k^2\frac{3T}{m_e}, ~~~    \omega_p^2=\frac{4\pi\alpha n_e}{m_e},  ~~~ \nu=\frac{\omega}{2\pi}, 
\end{equation}
where $n_e$ is the electron number density in the corona,  while $T$ is the temperature at the same altitude and $k$ is the wavenumber.
For example, the frequency $\nu=160~ {\rm MHz}$ considered in  \cite{Mondal-2020} will be emitted when $n_e\simeq 3.4 \cdot 10^{8}{\rm cm^{-3}}$. One should emphasize that the emission of radio waves generically occurs at an altitude  which is distinct  from the altitude where the AQN annihilation events occur, and where the energy is injected into the plasma. This is because the mean-free path $\lambda$ of the non-thermal electrons being injected into the plasma is very long $\lambda\sim 10^4$ km. Therefore, these electrons can travel a  very long distance before they transfer their energy to the radio wave, as we discuss in subsection \ref{non-thermal}. 

We propose that non-thermal electrons  are produced  by  anti-nuggets entering the solar corona, when the annihilation processes start. It is known that the number density of the non-thermal (suprathermal in terminology  \cite{Thejappa-1991}) electrons $n_{s}$   must be sufficiently large  $n_{s}/n_e\gtrsim  10^{-7}$ for the plasma instability to develop, in which  case   the radio waves will be generated \cite{Thejappa-1991}.  As the density $n_{s}/n_e$ approaches the threshold values at some specific frequencies, the intensity increases sharply, which we identify with the observed impulsive radio events. These threshold conditions may be satisfied randomly in space and time, depending on properties of the injected electrons  \cite{Thejappa-1991}. All these plasma properties  are well beyond the scope of this paper. However, we shall demonstrate that 
  the number density of the non-thermal electrons $n_{s}$  generated by the AQNs can easily be in proper range $n_{s}/n_e\gtrsim  10^{-7}$  for the plasma instability to develop.  To be  more specific, in next subsection \ref{non-thermal} we shall  argue that the ratio $n_s/n_e\sim 10^{-7}$ is always sufficiently large for the plasma instability to develop, which eventually generate the radio waves.   

Therefore, our proposal is that the AQN annihilations  (identified with nanoflares as explained in Section \ref{identification}) produce a large number of non-thermal electrons, which, in turn, generate the observed impulsive radio events  \cite{Mondal-2020} as a result of plasma instability. In the next subsections, we will support our proposal by estimating  a number of observables analyzed in  \cite{Mondal-2020}  . and show that our proposal is consistent with all observed data, including the frequency of appearance, the intensity radiation,  duration, spatial and wait time distributions,
to be discussed in next subsections \ref{rate}, \ref{non-thermal},  \ref{energetics}  as well as in Sections  \ref{wait_times}. 

\subsection{The event rate}\label{rate}

We are now in a position to interpret the radio emission data from  \cite{Mondal-2020} in terms of AQN annihilation events. The anti-nuggets start to loose their baryon charge, due to the annihilation,  in close vicinity of the transition region,  at an altitude of 2150 km ( see Fig. 5 in \cite{Raza:2018gpb} and also Fig. \ref{fig:ann_h} below).  However, the radio emission happens at much higher altitudes, as we explain in subsection \ref{non-thermal}.

\begin{figure*}
    \centering
    \includegraphics[width=0.8\linewidth]{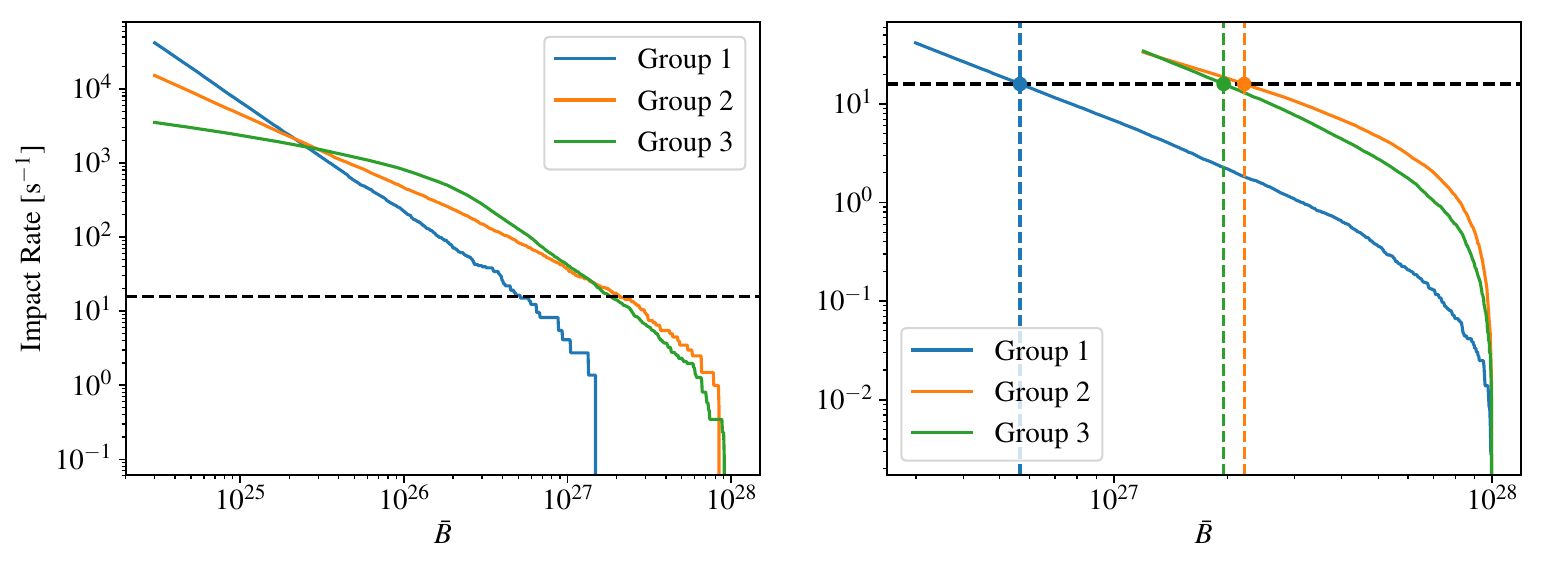}
    \caption{Left: the impact rate of AQNs with the size above $\bar{B}$ where $\bar{B}$ varies from $B_{\rm min}$ to $B_{\rm max}$ for different groups of AQNs. The horizontal black dashed line is the observed rate of radio events (\ref{eq:observation_rate}). Right: the result from the second-round simulation where we focus on large AQNs only. Again, the horizontal black dashed line is (\ref{eq:observation_rate}). The vertical dashed lines are the corresponding $\bar{B}$ for different groups. More details about the numerical simulations that lead to these two subfigures can be found in Appendix~\ref{appendix:simulations}.}
    \label{fig:impact_rate}
\end{figure*}

\begin{figure*}
    \centering
    \includegraphics[width=0.8\linewidth]{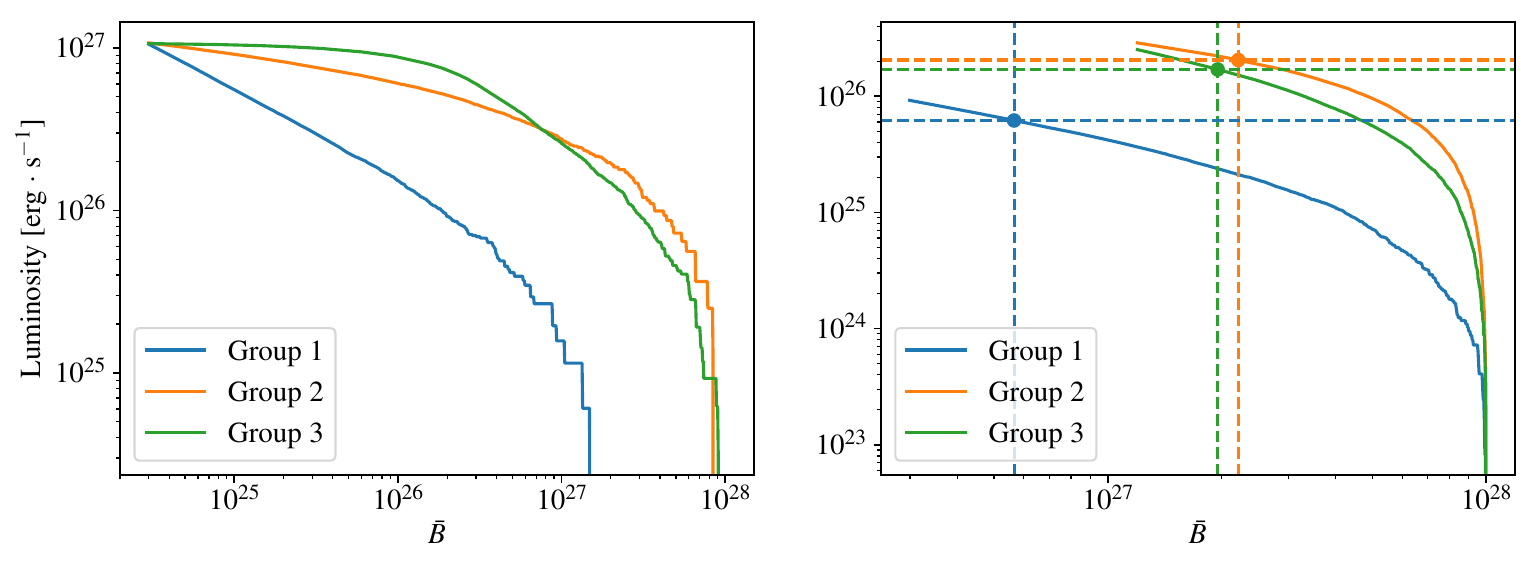}
    \caption{Left: the luminosity generated by the annihilation of AQNs with the size above $\bar{B}$ where $\bar{B}$ varies from $B_{\rm min}$ to $B_{\rm max}$ for different groups of AQNs. Right: the result from the second-round simulation where we focus on large AQNs only. The vertical dashed lines corresponds to the $\bar{B}$ determined by (\ref{eq:observation_rate}) in Fig.~\ref{fig:impact_rate}. More details about the numerical simulations that lead to these two subfigures can be found in Appendix~\ref{appendix:simulations}.}
    \label{fig:impact_luminosity}
\end{figure*}

In this subsection, we want to compare the maximum radio event rate (33,481 events  observed in the 132 MHz frequency band, during 70 minutes) to the expected rate of AQN annihilation events which are identified with nanoflares, and must be much more numerous (according to conventional solar physics modelling). Specific nanoflare models \cite{Pauluhn:2006ut,Bingert:2013} (expressed by  eq.(\ref{eq:groups}) in terms of the baryon charge $B$)  correspond to events rate  which is at least few orders of magnitude higher than the observed radio event rate, see Fig 8 in  \cite{Raza:2018gpb}. There is no contradiction here because it is likely that the dominant portion of the nanoflare events are too small to be resolved. This point has been mentioned  in items 2 and 3 in section  \ref{nanoflares} with a comment that  all  models must include small but frequent   events which had been extrapolated to sub-resolution region. Therefore, we interpret the low event rate at radio frequencies as the manifestation that
 only the strongest and the most energetic, but  relatively rare, AQN annihilation events can be resolved in radio bands. 
We define $\bar B$ as the minimum baryonic charge a nugget must have in order to generate a resolved radio impulse.  
 
 We can compute (in terms of $\bar{B}$)  the  event rate  for the energetic  AQNs  which  are powerful enough  to  generate the {\it resolved} radio impulses as recorded in   \cite{Mondal-2020}.
The corresponding impact rate can be computed in the same way as    Fig 8 from  \cite{Raza:2018gpb}, the only difference being that the lower bound is determined 
by $\bar{B}$ instead of $B_{\rm min}$, i.e.
\be
\label{event-rate}
\left(\frac{dN}{dt}\right)_{\bar{B}} \propto  \int_{\bar{B}}^{B_{\rm max}}\rm d B~ f(B).
\ee

Since the maximum number of detected radio events in \cite{Mondal-2020} is 33481 at the 132 MHZ band in 70 minutes, the event rate is
\begin{equation}
\label{eq:observation_rate}
\frac{dN_{\rm obs.}}{dt} \sim \frac{33481}{70{\rm~minutes} \times 1/2} \sim 16{~\rm s^{-1}}.
\end{equation}
The factor $1/2$ accounts for the fact that only half of the Sun's whole surface is visible. 

By equalizing  (\ref{eq:observation_rate}) and (\ref{event-rate}) we can estimate the parameter $\bar{B}$  when sufficiently large radio events originate from large nuggets with $B\gtrsim \bar{B}$ \footnote{This estimate does not include the possibility of ``clustering" events with very short time scale discussed
in Section \ref{clustering}.}. The results are presented on Fig. \ref{fig:impact_rate}. It is the intersection of the black dashed line (\ref{eq:observation_rate}) and the simulated line of each group given by eq.(\ref{eq:groups}). The intersections are shown in the right subfigure, and the corresponding $\bar{B}$ are respectively $5.65\times 10^{26}$, $2.21\times 10^{27}$,  and $1.95\times 10^{27}$ for the three groups.
We expect that only AQNs with masses greater  than $\bar{B}$ are sufficiently energetic to generate the observable impulsive radio events.

The parameter $\bar{B}$  obviously depends on the size distribution models  listed in (\ref{eq:groups}), it corresponds to a detection limit and should not be treated as a fundamental parameter of the theory. An instrument with different resolution and/or sensitivity will affect the radio events selection criteria and therefore change the value of $\bar{B}$, in which case some events from the continuum spectrum would be considered as impulsive events\footnote{It is known that the continuum contribution in the radio emissions is similar in magnitude to the impulses events as we discuss in subsection \ref{energetics}. Some of the  events from continuum  could be treated in future as impulsive events if a  better resolution instrument is available. However, this does not drastically modify our estimate for    $\bar{B}$.}.
  
Our next task is to estimate the total luminosity $L^{\odot}_{\bar{B}} $ released as a result of the  complete annihilation of the large nuggets with $B\gtrsim \bar{B}$ . The calculation is similar to the estimation done for Fig 10 of \cite{Raza:2018gpb}, the only difference is that the lower bound is determined 
by $\bar{B}$ rather than $B_{\rm min}$, i.e. 
\be
\label{energy-rate}
L^{\odot}_{\bar{B}} \propto  \int_{\bar{B}}^{B_{\rm max}}\rm d B~ B^{\frac{2}{3}} f(B).  
\ee
 The results for the models listed in   (\ref{eq:groups}) are presented on Fig. \ref{fig:impact_luminosity}. The corresponding $L^{\odot}_{\bar{B}}$  assume the following values:  $6.17\times10^{25} {\rm~erg\cdot s^{-1}}$, $2.05\times10^{26} {\rm~erg\cdot s^{-1}}$ and $1.70\times10^{26} {\rm~erg\cdot s^{-1}}$, which are approximately an order of magnitude smaller than the luminosity released by all the AQNs annihilation. This implies that only about 10\%   of the total the AQN- induced luminosity comes from the large nuggets with $B\gtrsim \bar{B}$, which are the same AQNs assumed to produce the resolved radio events in \cite{Mondal-2020}. Our estimates show that while the strong events with $B\gtrsim \bar{B}$ are very rare with an impact rate approximately 3 orders of magnitude smaller than all AQN annihilation events, their contribution to the luminosity is suppressed only by one order of magnitude. This is, of course, due to the factor $B^{\frac{2}{3}}$ in the expression for the luminosity  (\ref{energy-rate}).

The energy flux $\Phi^{\odot}_{\bar{B}}$, observed on Earth, coming from these large nuggets with $B\gtrsim \bar{B}$   is estimated as 
\be
\label{flux}
\Phi^{\odot}_{\bar{B}}\simeq \frac{L^{\odot}_{\bar{B}}}{4\pi (AU)^2}   \simeq      (1.8-6) \cdot 10^{-2}\frac{\rm erg}{\rm cm^2~ s}   , 
\ee
where we used the range of numerical values for $L^{\odot}_{\bar{B}}$ estimated above.
In the following we will establish the physical connection between the energy flux (\ref{flux}) generated by large nuggets with $B\gtrsim \bar{B}$   and the flux observed in radio frequency bands
  observed in   \cite{Mondal-2020}. In order to make this connection we have to estimate what fraction of the huge amount of energy due to the AQN  annihilation is transferred to the tiny portion in the form of radio waves.
   To compute this efficiency we need to estimate the relative density of the non-thermal electrons which 
   will be produced as a result of the AQN annihilation events. 
    The estimation of this efficiency   is the topic of the next subsection.

\subsection{Non-thermal electrons}\label{non-thermal}
The starting point for our analysis is the number of annihilation events per unit length while the AQN propagates through the ionized corona environment:
\be
\label{rate-annihilation}
\frac{dN}{dl}\simeq \pi R_{\rm eff}^2 n_{\rm p},
\ee 
where $n_{\rm p}$ is the baryon number density of the corona (mostly protons) and the effective radius $R_{\rm eff}$ of the AQNs can be interpreted as the effective size of the nuggets due to the ionization characterized by  the nugget's charge $Q$ as explained in  \cite{Raza:2018gpb}. The enhancement of the interaction range $R_{\rm eff}$ due to the long range Coulomb force is given by (see \cite{Raza:2018gpb} for the details): 
 \be
 \label{epsilon}
 \left(\frac{R_{\rm eff}}{R}\right) = \epsilon_1 \epsilon_2, ~~ \epsilon_1 \equiv \sqrt[]{\frac{8 (m_e T_P) R^2}{\pi}}, ~~
 \epsilon_2\equiv \left(\frac{T_I}{T_P}\right)^{\frac{3}{2}},  ~~~~
 \ee
where $T_I$  is  the internal temperature of the AQN and $T_P$ is the plasma temperature in the corona.
The estimation of the internal thermal temperature $T_I$  is a highly nontrivial and complicated problem which requires an understanding of how the heat, due to the friction and the annihilation events, will be transferred to the surrounding plasma from a body moving with supersonic speed with Mach number $M\equiv v/c_s>1$. 

It is known that the supersonic motion will  generate shock waves and turbulence.  It is also  known   that a shock wave leads to a discontinuity in velocity, density and temperature due to the  large Mach numbers $M\gg 1$. It has been argued in \cite{Zhitnitsky:2018mav,Raza:2018gpb} that, for a normal shock, the jump in temperature is given by the  Rankine–Hugoniot condition:
\be
 \label{shock1}
  \frac{T_I}{T_P}\simeq M^2\cdot \frac{2\gamma(\gamma-1) }{(\gamma+1)^2}\gg 1, ~~~ \gamma\simeq 5/3,
 \ee
and, as a result, all the electrons from the plasma which are on the AQN path within distance $R_{\rm eff}$ will be affected.  To be more precise  these electrons will experience elastic scattering by receiving the  extra kinetic energy $\Delta E$ which lies in the window $\Delta E\in (T_P, T_I)$. It is precisely these non-thermal electrons which will subsequently interact with the plasma and be the source of the plasma instability. These  non-thermal electrons will transfer their energy to  the emission of radio waves with frequency $\nu$ as explained  at the end of Section \ref{mechanism}.

We are now in position to estimate the parameter $\eta$ defined as the ratio  between   the energy transferred (per unit length $l$) to the radio waves and the  total  energy produced by a single AQN (per unit length $l$) as a result of the annihilation process:
\be
\label{eta}
\eta\approx \frac{(\Delta E)\cdot [\pi R_{eff}^2 n_{\rm e}]}{(2m_pc^2) \cdot [\pi R_{eff}^2 n_{\rm p}]}\approx  \frac{\Delta E}{2m_pc^2}\sim 10^{-7},
\ee
where the denominator accounts for the total energy due to the annihilation events with rate (\ref{rate-annihilation}) and the numerator accounts for the kinetic energy received by affected electrons. In our estimate of (\ref{eta}),  we assume an approximate local neutrality such that $n_e\approx n_p$. Furthermore, to be on the conservative side,  we also assume  that  $\Delta E \approx 2\cdot 10^2 ~{\rm eV}$, such that $\Delta E$ only slightly exceeds the plasma temperature $\approx T_P$ at high altitudes of order $10^4$ km, where radio emission occurs. Finally, we also assume  that the dominant portion of the $\Delta E $ will be eventually released in the form of radio waves.   It is very likely that there are  few  missing numerical factors of order one on the right hand side in eq. (\ref{eta}) as   our assumptions formulated above are only approximations. However, we believe that  (\ref{eta}) gives  a correct order of magnitude estimate for the energy efficiency  transfer  ratio $\eta$. We provide a few numerical estimates in next subsection \ref{energetics}   suggesting that (\ref{eta}) is very reasonable and consistent with observed intensities in radio bands  \cite{Mondal-2020}.   

The next step is the estimation of $n_{s}/n_e$, which must be sufficiently large $n_{s}/n_e\gtrsim  10^{-7}$ for the plasma instability to develop \cite{Thejappa-1991} (see section \ref{mechanism}).  As we shall see now, the proposed mechanism indeed satisfies this requirement. 
We start with the expression of the total number of  electrons $\Delta N_e$ to be affected while the AQN travels over a distance $l $:
\be
\label{N_e1}
 \Delta N_e\sim  (\pi R_{eff}^2   l ) \cdot n_{\rm e}(h) , ~~~~ l\simeq v_{\rm AQN}\Delta t,
\ee 
where $n_{\rm e}(h)$ is the electron number density at the altitude $h\simeq 2000$ km where annihilation events become efficient \cite{Raza:2018gpb}.
These affected electrons will receive an extra energy $\Delta E$ and extra momentum $m_e{v}_{\perp}$  with very large velocity component ${v}_{\perp}$ perpendicular to the nugget's path as the shock front due to $M\gg 1$ has a form of a cylinder along the AQN path. A large portion of  the AQN's trajectories can be viewed  as an almost horizontal path with relatively small incident angles toward the Sun (skim trajectories). These non-thermal electrons will have a component ${v}_{\perp}$ perpendicular to the nugget's path and travel unperturbed up to a distance of the order of the mean free path $\lambda\sim 10^4~ {\rm km}$ (to be estimated below).

After a time $\Delta t$, the same non-thermal electrons $ \Delta N_e$ will have spread over a distance $r $ from the AQN's path, estimated as follows:
\be
\label{N_e2}
 \Delta N_e\sim  (2\pi r\Delta r    l ) \cdot n_{\rm s} (r), 
 \ee 
where $\Delta r$ is the width of the shock front measured at
distance $r$. For a non-thermal electron traveling away from the AQN path with perpendicular velocity ${v}_{\perp}$, the distance $r$ is given by:
\be
\label{N_e3}
  r\sim   {v}_{\perp}\Delta t, ~~~{v}_{\perp}\simeq \sqrt{\frac{2\Delta E}{m_e}} \simeq   10^4 \sqrt{\frac{\Delta E}{2\cdot 10^2 {\rm~eV}}} \rm \frac{km}{s} .
\ee 
Equalizing  (\ref{N_e1}) and  (\ref{N_e2})    we arrive to the following estimate for the ratio $n_{s}/n_e$:
\be
\label{n_s1}
\left[\frac{n_s (r)}{n_e(h)}\right]\simeq \left(\frac{R_{eff}^2}{ r\Delta r}\right)  , ~~~ r\lesssim  \lambda .
\ee
 The expression (\ref{n_s1}) holds as long as $r\lesssim  \lambda$. For larger distances  $r\gtrsim \lambda$   the non-thermal electrons will eventually thermalize and loose their ability to generate a plasma instability. One should emphasize that $n_s (r)$ entering (\ref{n_s1}) is taken at the distance $r$ from the AQN path, while 
 $n_e(h)$ is taken in the vicinity of the path, i.e. at $r\approx 0$.  
 
 We are interested in this ratio when both components 
 are computed at the same location and we now have to check if it is larger than $10^{-7}$, the requirement to generate the plasma instability. The relevant configuration for our study corresponds to non-thermal electrons moving upward\footnote{the radio waves emitted at altitudes below $h$ will have much higher frequencies 
 than considered in the present work, and shall not be discussed here.}. In this case the relation 
 (\ref{n_s1}) assumes the form
 \be
\label{n_s3}
\left[\frac{n_s (r)}{n_e(r)}\right]\simeq \frac{1}{2}\left[\frac{n_e (h)}{n_e(r+h)}\right] \cdot \left(\frac{R_{eff}^2}{ r\Delta r}\right)  , ~~~ r\lesssim  \lambda , 
\ee
 where the factor 1/2 accounts for upward moving electrons and ${n_e (r)}\equiv {n_e(r+h)}$ is the electron density computed at distance $\sim r$ above the AQN's path (which is localized  at an altitude of $h\simeq 2000$ km).
 
 The expression (\ref{n_s3}) has a conventional form for a cylindrical geometry with the expected suppression factor $r^{-1}$ at large distances and constant value for $\Delta r$.
 However, it is known that the width of the shock $\Delta r$ also growths  with 
 time\footnote{Such scaling is known to occur, for example, when the meteoroids  propagate in the Earth's atmosphere when the cylindrical symmetry is also realized. We refer to \cite{Budker:2020mqk} (with large list of references on the original literature devoted to this topic) where this scaling specific for the cylindrical  geometry has been used in the context of the AQN propagation   in Earth's atmosphere.}  as $\Delta r\propto \sqrt{t}\propto \sqrt{rR_{eff}}$. Therefore, we expect that a proper scaling at large $r$ assumes the form:
 \be
\label{n_s4}
\left[\frac{n_s (r)}{n_e(r)}\right]\sim \frac{1}{2}\left[\frac{n_e (h)}{n_e(r+h)}\right] \cdot \left(\frac{R_{eff}}{ r }\right)^{\frac{3}{2}}   , ~~~ r\lesssim  \lambda , 
\ee
We will calculate this ratio for large nuggets with $B\gtrsim \bar{B}$ which are capable of generating the resolved  radio signals. 
  Using our previous parameters estimates for $\epsilon_1$ and $\epsilon_2$ from Section IV.C of
  \cite{Raza:2018gpb} and using the electron number density in Table 26 of \cite{2008ApJS..175..229A} , we arrive at the estimate 
  \be
\label{n_s2}
\left[\frac{n_s (r)}{n_e(r)}\right]\gtrsim 10^{-7} ~~{\rm for }~~  r \sim 10^4 ~{\rm km}.
\ee
 The condition (\ref{n_s2}) implies that  $n_{s}/n_e$ is indeed sufficiently large for the plasma instability to develop \cite{Thejappa-1991} on distances of order $ r\sim    10^4 ~{\rm km}$   from the nugget's path. This implies that the non-thermal electrons can propagate upward to very large distances  before they transfer their energy to the radio waves at much higher altitudes, of order $(h+r)$. 
 The scale $ r\sim    10^4 ~{\rm km}$ assumes the same order of magnitude value as the    mean free path $\lambda$,
  which  at altitude $h\simeq 10^4 $ km can be estimated as follows:
 \be
 \label{lambda}
 \lambda^{-1}\simeq \sigma n_p, ~~ \sigma\simeq \frac{\alpha^2}{(\Delta E)^2} , ~~~\lambda \sim   10^4 ~{\rm km},
 \ee
 where $(\Delta E)\approx 2\cdot 10^2$ eV is the typical kinetic energy of the  non-thermal electrons at the moment of emission. 
 
   One should emphasize that the estimation given above assumes a constant density  $n_p$ along the electron's path. This is clearly not the case for the upward moving non-thermal electrons. One can define an effective mean free path $ {\lambda}_{\rm eff}^{-1}(h)$ as follows\footnote{We use $h_{0}\simeq 2150$~km precisely because AQNs start to annihilate at this altitude (shown in Fig.~\ref{fig:ann_h}) which is the start of the transition region as the density drastically increases (see Table 26 or Fig.~8 of \cite{2008ApJS..175..229A}).}
 \be
     \label{lambda_eff}
      {\lambda}_{\rm eff}^{-1}(h)\equiv\int_{h_0}^{h} \frac{dh'  \sigma n_p(h')}{(h-h_0)},  ~~~~ h_0\simeq 2150 {\rm km} ,
     \ee
 which accounts for the density variation with altitude. It reduces to the canonical definition (\ref{lambda}) when $n_p$ is a constant along the electron's path. This definition of the effective mean free path in the context of the present proposal is very convenient as it explicitly shows at what altitude most of the energy will be thermalized, and what portion of the energy can be released in form of the radio waves.
 
\begin{figure}
    \centering
    \includegraphics[width=1\linewidth]{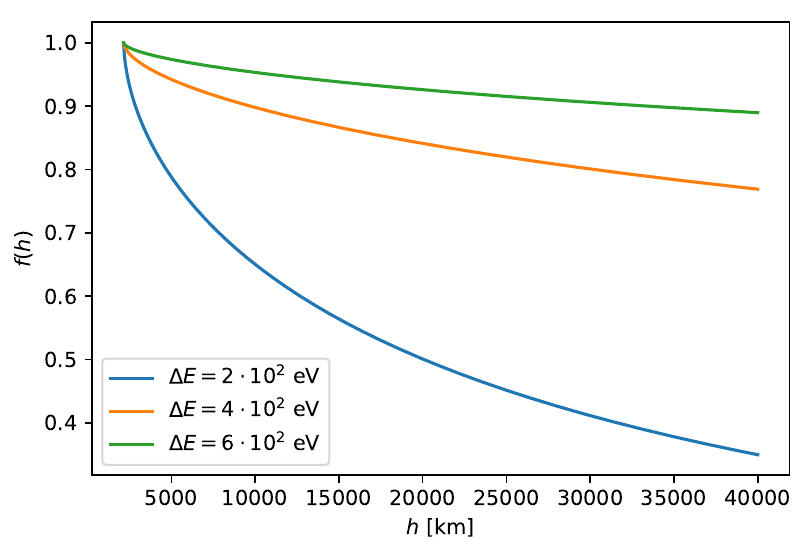}
    \caption{Suppression factor $f(h)$ defined by eq. (\ref{survival}).
    This factor describes the remaining portion of the non-thermal electrons at altitude $h$. The blue line corresponds to the initial kinetic energy $\Delta E\approx 2\cdot 10^2$ eV which has been used in all our estimates through the text. For illustrative purposes we also presented the same  suppression factor $f(h)$ for other values of parameter $\Delta E$. Suppression factor becomes essential for $h\gtrsim 4\cdot 10^4$ km corresponding to low frequency emission as one can see from Fig.\ref{fig:frequency}. In computing (\ref{survival}), we have used $n_{p}(h)\approx n_{e}(h)$ above $h_{0}$ where the profile of $n_{e}(h)$ is from \cite{2008ApJS..175..229A} (the solar profiles needed in the numerical computations in this work are all from \cite{2008ApJS..175..229A}).}
    \label{fig:survival}
\end{figure}

 To be more precise, the portion $f(h)$ of the non-thermal electrons which survives at altitude $h$  can be estimated as follows
 \be
 \label{survival}
 f(h)=\exp  \left(-\int_{h_0}^{h}\frac{dh'}{\lambda_{\rm eff}(h')}\right),
 \ee
 where mean free path $\lambda_{\rm eff}(h)$ at  altitude  $h$ is defined by eq. (\ref{lambda_eff}). The behaviour for $f(h)$ as a function of the altitude $h$ is shown on Fig. \ref{fig:survival} by blue line for initial kinetic energy of the non-thermal electrons $\Delta E\approx 2\cdot 10^2$ eV. This value  for $\Delta E$  has been used in all our estimates through the text.
 
\begin{figure}
    \centering
    \includegraphics[width=1\linewidth]{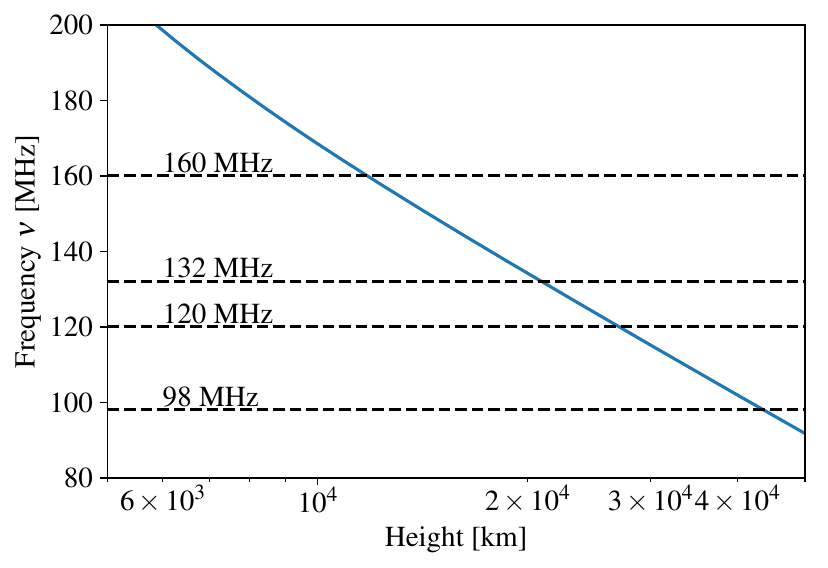}
    \caption{Frequency of the emission $\nu=\omega/2\pi \approx \omega_p/2\pi$ (i.e., eq. (\ref{eq:omega})) as a function of height. Radio emission occurs at the altitudes above  $10^4$  km while the dominant portion of the AQN annihilation events occur at lower altitudes $h< 2150$~ km as shown on Fig.\ref{fig:ann_h}.}
    \label{fig:frequency}
\end{figure}
The most important remark here is that the suppression factor $f(h)$
is very modest for altitudes where high frequency waves are emitted, see Fig.\ref{fig:frequency}. We emphasize that, in this parameters range, the density of the non-thermal electrons remains sufficiently large  to satisfy the crucial  condition (\ref{n_s2})  for the plasma instability to develop \cite{Thejappa-1991}. Therefore,  the dominant portion of the non-thermal electron's energy will be released in form of the radio waves.  

 At the same time the suppression   becomes essential for higher altitudes where low frequency waves are emitted.  
 At higher altitudes the suppression factor $f(h)$ plays the dominant role and non-thermal electrons loose their energy to thermalization. The density of the non-thermal electrons  is  insufficient    to satisfy the crucial  condition (\ref{n_s2})  for the plasma instability to develop \cite{Thejappa-1991}.  At this point the    radio emission  stops completely. One should emphasize that such a sharp cutoff for the radio emission at lower frequencies is very unique and specific prediction of the proposed mechanism.

 One should also mention that   the density $n_p$ drastically increases at slightly lower altitudes $h\lesssim 2000$~ km (in comparison with $h\simeq 2150$ km), such that the mean-free path $\lambda_{\rm eff}$ decreases correspondingly, and the condition (\ref{n_s2}) breaks down.  Therefore, the non-thermal electrons emitted at  $h\lesssim 2000$~ km  cannot propagate to very high altitudes $\sim 10^4$ km where radio emission occurs.

 \subsection{Radio flux intensity}\label{energetics}
 \begin{figure*}
    \centering
    \includegraphics[width=0.8\linewidth]{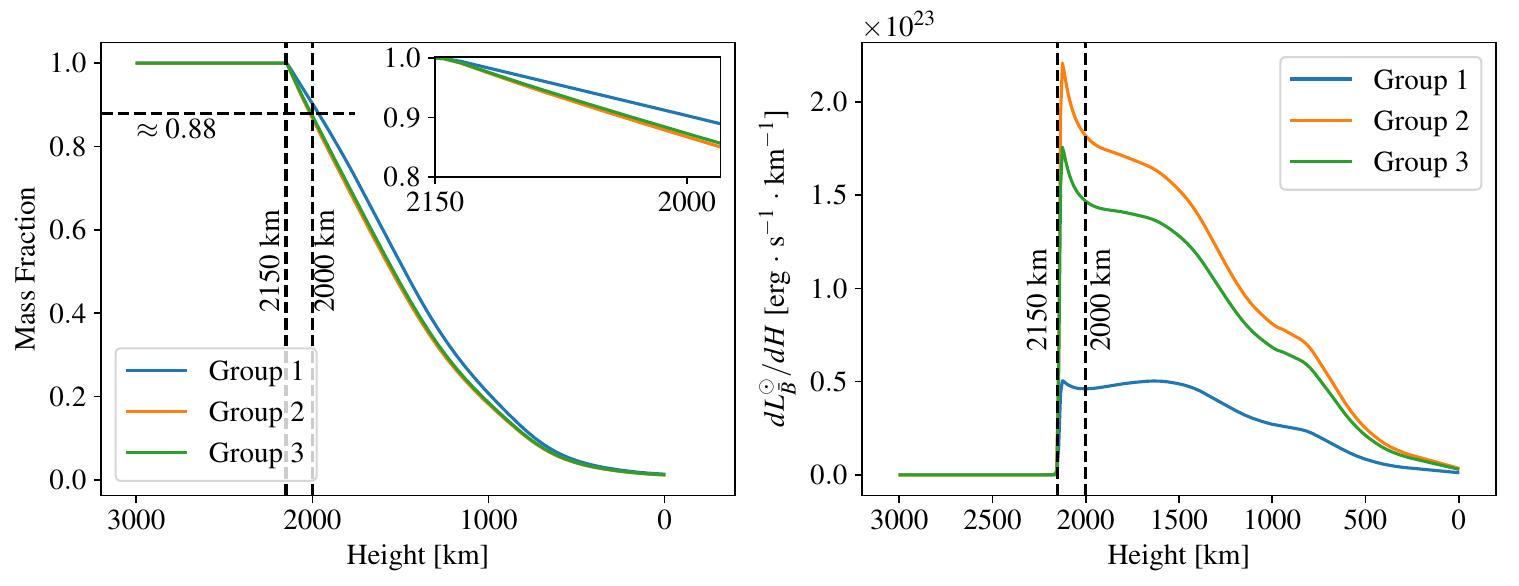}
    \caption{Left: Mass fraction $1-{\Delta B}/{B}$ being annihilated as a function of the altitude. This is plotted by taking the average of the mass loss profiles of the AQNs above $\bar{B}$ (i.e. the AQNs that will generate radio emissions) where $\bar{B}$ has been determined by (\ref{eq:observation_rate}). We see that the AQNs start to annihilate at about $2150$~km. Right: Luminosity per unit length as a function of the altitude where the energy is converted from the mass loss according to (\ref{units}). This is plotted also by taking the average of the AQNs above $\bar{B}$, then multiplied by the impact rate of these large AQNs.}
    \label{fig:ann_h}
\end{figure*}

In this subsection we  estimate  the portion of the AQN-induced energy flux  which is  transferred to the radio waves $\Phi^{\rm radio}$. We express $\Phi^{\rm radio}$ in terms of the energy flux emitted by the nuggets as radio waves:
 \be
 \label{flux-radio}
&&\Phi^{\rm radio}\simeq  \Phi^{\odot}_{\bar{B}}\cdot \eta   \left(\frac{\Delta B}{B}\right) \\
&\simeq& (0.6-6)\cdot  10^{-10}  \frac{\rm erg}{\rm cm^2~ s},~~~~~~~ (\rm theoretical ~prediction) \nonumber
 \ee
where the first factor  $\Phi^{\odot}_{\bar{B}}$, given by (\ref{flux}), reflects the contribution of the large nuggets with $B\gtrsim \bar{B}$ to the total AQN-induced luminosity. The factor $\eta$ is given by (\ref{eta}) and represents the portion of the energy transferred to the radio frequency bands through the non-thermal electrons leading to the plasma instability. Finally, 
the factor  ${\Delta B}/{B}\sim (0.3-1)\cdot 10^{-1}$  describes a typical portion of the baryon charge  annihilated in the altitude range  (2000-2150) km. This is precisely the region where the AQN annihilation events effectively start and where the interaction of the AQNs with surrounding plasma  produce  the non-thermal electrons which eventually generate the radio waves. The  Monte-Carlo simulations for $ {\Delta B}/{B}$ are presented on Fig. \ref{fig:ann_h}. One can see that the dominant portion of the annihilation events occur at the lower altitudes $h\lesssim 2000$ km. However, the mean free path $\lambda$ at lower altitudes of the affected electrons is too short as our estimations (\ref{lambda}) suggest. Therefore, the affected electrons from altitudes $h\lesssim 2000$ km cannot reach higher altitudes where the radio waves are generated. This is precisely  the source of the suppression expressed in the ratio ${\Delta B}/{B}\ll 1$.

We can now compare our estimate (\ref{flux-radio}) to the observed  intensities measured in radio frequency bands by \cite{Mondal-2020}:
\be
\label{flux-observed}
\frac{d \Phi^{\rm radio} }{d \omega} (\rm 160~ MHz)\simeq 6~ SFU, ~~~   \Delta \omega=2.56~ MHz  \nonumber\\
\frac{d \Phi^{\rm radio} }{d \omega} (\rm 120~ MHz)\simeq 3~ SFU, ~~~   \Delta \omega=2.56~ MHz  
\ee
where 
\be
\rm SFU\equiv  10^4 Jy=10^{-19}  \frac{\rm erg}{\rm Hz~cm^2~ s}.
\ee
The observations \cite{Mondal-2020} were done in twelve frequency bands from $\rm 80~MHz$ to $\rm 240~ MHz$  with 
$ \rm \Delta \omega=2.56~ MHz$ bandwidth each. It is known \cite{Sharma_2018,oberoi_2017} that the radio emission occurs in the entire energy band $\sim \rm (0-200) ~MHz$, and not specifically in one of the 12 frequency narrow bands. It is also known \cite{Sharma_2018,oberoi_2017} that the contributions from  continuum and impulsive fluxes are approximately the same in all frequency bands.   Therefore we estimate  the total   intensity  in radio bands 
by multiplying  (\ref{flux-observed}) with $\rm  \sim 200~ MHz$ to account for the entire radio emission associated with short impulsive events as well as the continuum:
\be
\label{flux-total}
\Phi^{\rm radio}_{\rm total}  \simeq (0.6-1.2) \cdot 10^{-10}    \frac{\rm erg}{\rm cm^2~ s} ~~~ (\rm observation).~~~
 \ee
Despite the fact that our calculation involves various steps and approximations, the total measured flux (\ref{flux-total}) is consistent with  our order of magnitude estimate (\ref{flux-radio}). 
We consider this as   a highly non-trivial consistency check for our proposal as it includes a number of very different elements which were studied previously for a completely different purpose in a different context.  

We conclude this section with few important remarks.
The occurrence probability shown on Fig 4 in \cite{Mondal-2020}   suggests that 
the power-law index $\alpha$ is always large, with $\alpha>2$.  As explained in the text we cannot predict this index theoretically, but all the nanoflare models used in our studies as expressed by eq. (\ref{eq:groups}) are consistent with the observed power-law index $\alpha$ because the nuggets generating the resolved radio impulses  must be  sufficiently large with  $B> \bar{B}$,   in which case the  index $\alpha$ is always large (index $\alpha=1.2$ for one of the model from  (\ref{eq:groups}) describes the distribution of small nuggets with $B< \bar{B}$ which do not produce the resolved radio signals). 

The basic picture for the radio emission   advocated here is that one and the same AQN may generate   the emissions in different frequency bands because the non-thermal electrons produced by the AQN and moving  in upward direction   can emit the radio waves at different altitudes with different plasma frequencies as long as non-thermal electron density is sufficiently high and satisfies the condition (\ref{n_s2}). As an illustration, we show the frequency of emission (\ref{eq:omega})  as a function of height on Fig.\ref{fig:frequency}.   
In this example, all the radio emissions must be correlated with in time over seconds, which is considerably shorter than the typical mass loss time scale which is about  10- 20 seconds, see Fig 5, 6  in   \cite{Raza:2018gpb}.

 This generic picture also suggests that the emission at higher frequencies $\nu$ must be more intense due to a number of reasons.
First, the upward moving non-thermal electrons are much more numerous
at lower altitude (corresponding to  higher  $\nu$) in comparison with higher altitudes (corresponding to   lower $\nu$) because $n_s/n_e$ ratio scales as $r^{-3/2}$. When this scaling reaches a ratio below the required rate (\ref{n_s2}) the radio wave emission cannot occur as the density of the non-thermal electrons is not sufficient for the plasma instability to develop \cite{Thejappa-1991}. Furthermore,   the effective mean free path  determined by  (\ref{lambda_eff}) 
essentially determines  the highest altitudes where non-thermal electrons may  reach, see Fig.\ref{fig:survival}.  After this height the non-thermal electrons will thermalize  
and cannot be the source of the radio waves.

Secondly, according to (\ref{flux-radio})  the lower the altitude, the higher the annihilation rate.    This is because the portion of the annihilated baryon charge $\Delta B/B$   drastically increases  when altitude decreases, see Fig. \ref{fig:ann_h}.  When the frequency of the radio emission  becomes too high, the radiation becomes a subject of absorption too strong to be detectable  above the quiet Sun background. Such suppression with higher frequency radiation has indeed been observed for frequencies $\nu\gtrsim 240~ {\rm MHz}$, see \cite{oberoi_2017}.

The same line of arguments may also explain the observed huge difference between the number of observed events (4748) at smallest frequency band (98  MHz) in comparison to the rate at larger frequency bands  where the recorded number of events is almost  one order of magnitude higher \cite{Mondal-2020}.    These arguments suggest that counting rate at even lower frequencies (such 80 and 89 MHz bands recorded by MWA)
should be even lower than 4748 events  recorded at 98  MHz  \cite{Mondal-2020}.

\section{The AQN model. Wait time distribution. }\label{wait_times}
The goal here is to understand the wait time distribution reported by \cite{Mondal-2020}.
The main observation was that the impulsive events are non- Poissonian in nature.
This non- Poissonian feature is shown on Fig 7 of \cite{Mondal-2020} where the occurrence probability  at small wait times (below 10 seconds) is linearly growing instead of approaching a constant, which is what is expected for a Poissonian distribution. 

We shall argue below that, in the AQN model, such a behaviour could be explained by the presence of ``effective" clustering of events when one and the same AQN in flight may generate a cascade of seemingly independent events on short time scales. 
These events however, are not truly independent, as they result, in fact,  from one and the same AQN when  the typical  mass loss time  is measured in 10-20 seconds, see Fig. 6 in  \cite{Raza:2018gpb}. Few short radio pulses on scales of few seconds could be easily generated during this long flight time. Such ``clustering" will violate the assumption of the Poissonian distribution of independent events.

In what follows we develop an approach which can incorporate such ``clustering" at small time scales, while the distribution remains Poissonian at larger time scales, i.e. the time scale of distinct AQNs entering the Corona. The corresponding approach is known as a non-stationary Poissonian process which results in Bayesian statistics, which is the topic of the next subsection. 
 
 \subsection{Non- Poissonian processes. Overview.}
 We start with an overview of the  non- Poissonian processes and outline the conventional technique to describe them, as given in \cite{Wheatland_1998, Aschwanden_2010,Li_2014}.  
 In case of a conventional random stationary Poissonian process, the waiting time distribution $P (\Delta t)$ is expressed as an exponential distribution:
 \be
 \label{Poissonian}
 P (\Delta t)=\lambda e^{-\lambda \Delta t}, ~~~~ \int P (\Delta t) d \Delta t=1, 
 \ee 
where $\lambda$ in this section is mean event occurrence rate. For a constant $\lambda$, this distribution describes a stationary 
Poissonian process. When $\lambda(t)$ depends on time, one can generalize (\ref{Poissonian}) and introduce the probability function of waiting times which becomes itself a function of time \cite{Wheatland_1998}:
\be
 \label{P(t)}
 P (t, \Delta t)=\lambda (t+\Delta t) \exp{\left[-\int_t^{t+\Delta t}\lambda(t') dt' \right]}.
 \ee 
If observations of a non-stationary Poisson process are made during a time interval $[0, T]$, then the distribution of waiting times $P (\Delta t)$ will be, weighted by the number of events $\lambda(t) dt$ in each time interval $(t, t+dt)$, given by:
\be
\label{P(t)1}
 P (\Delta t)=\frac{1}{N}\int_0^T\lambda(t)P (t, \Delta t) dt, ~~~ N=\int_0^T\lambda(t) dt. ~~~
\ee
If $\lambda$ varies adiabatically one can subdivide non-stationary Poisson processes  into piecewise stationary Poisson processes (Bayesian blocks), take the continuum limit and represent the distribution of waiting times as follows \cite{Wheatland_1998, Aschwanden_2010,Li_2014}:
\be
\label{P(t)2}
 P (\Delta t)= \frac{\int_0^T\lambda^2(t) e^{-[\lambda(t) \Delta t]}dt}{\int_0^T\lambda(t)dt}.
\ee
One can check that the expression (\ref{P(t)2}) reduces to its original Poissonian expression (\ref{Poissonian}) when $\lambda$ is time independent.

It is convenient to introduce $f(\lambda)$ which describes the adiabatic changes of $\lambda$ as follows:
\be
\label{f}
f(\lambda)\equiv \frac{1}{T} \frac{dt(\lambda)}{d \lambda}, ~~~~~ f(\lambda)d\lambda=\frac{dt}{T}, ~~ \int d\lambda f(\lambda)=1. ~~~~
\ee
In terms of $f(\lambda)$ the distribution of waiting times (\ref{P(t)2})  assumes the form
\be
\label{P(t)3}
 P (\Delta t)= \frac{\int_0^{\infty}\lambda^2f(\lambda) e^{-[\lambda \Delta t]}d\lambda}{\int_0^{\infty}\lambda f(\lambda)d\lambda}.
\ee
The stationary 
Poissonian distribution corresponds to $f(\lambda)=\delta(\lambda-\lambda_0)$ such that the distribution of waiting times (\ref{P(t)3}) reduces to the original expression (\ref{Poissonian}) with constant $\lambda_0$ as it should.

\subsection{AQN induced clustering events}\label{clustering} 

We are now in position to describe the physics  of ``effective" clustering events using non-stationary Poisson distribution framework (\ref{P(t)3})  as outlined above. 
As previously mentioned several, short radio pulses on scales of few seconds could be easily generated during a single AQN ``relativly'' long flight time of the order of  10-20 seconds (see Fig. 6 in  \cite{Raza:2018gpb}).

With this picture in mind, we introduce the following  $\lambda(t)$ dependence to describe non-stationary Poisson processes. At long time scales $ t> t_0$ we keep  the constant $\lambda_0$ corresponding to the stationary Poisson distribution:  
\be
\label{lambda1}
 \lambda=\lambda_0 ~~   f(\lambda)\sim\delta(\lambda-\lambda_0), ~~~{\rm for} ~~~ t> t_0,~~~~
\ee 
while for shorter time scales $t<t_0$ we parameterize $f(\lambda)$ as follows:
\be
\label{lambda2}
  f(\lambda)=c\lambda^{\beta}, ~~~\lambda=\lambda_0\left[\frac{t}{t_0}\right]^{\frac{1}{\beta+1}}~~~ {\rm for} ~~~ t<t_0,~~~~
\ee 
where $\beta, \lambda_0$ and $t_0$ parameters should be fitted to match the observational signal distribution.
 The parameterization for non-stationary Poisson processes (\ref{lambda2}) is a generic power law behaviour which satisfies the condition $\lambda (t\rightarrow 0)\rightarrow 0$   when  $t\rightarrow 0$.   
It has been used previously \cite{Wheatland_1998, Aschwanden_2010, Li_2014} for many different systems, including  the solar flares\footnote{In particular, in \cite{Li_2014} a more general expression for $f(\lambda)=c\lambda^{\beta}\exp(-\gamma\lambda)$ was considered which also includes the exponential tail $\exp(-\gamma\lambda)$. We do not include this exponential factor as it simply shifts the definition for $\Delta t\rightarrow (\Delta t+\gamma)$ as one can see from eq. (\ref{P(t)3}).}. In comparison with previous studies we consider the superposition of two terms (\ref{lambda1}) and (\ref{lambda2}) which allows us to quantitatively characterize (by taking an appropriate limit) the level of  non-stationary Poisson processes and the extend of deviation from the stationary Poisson distribution. As we shall argue below, the non-stationary Poisson processes play the dominant role in  our studies, which is the main claim  of the present section. 

 We start by   explaining  the  physical   meaning of the parameters entering  (\ref{lambda1}) and (\ref{lambda2}). 
 As we discuss below the  $t_0$ will enter the observables in form of the dimensionless parameter $(t_0/T)$. The physical meaning of this parameter $(t_0/T)$ is clear:
it determines the time-portion of the clustering events.
In case when $(t_0/T)\ll 1$ the clustering events play a very minor role, while for $(t_0/T)\sim  1$ the clustering events become essential. 
In the limit $t_0/T\rightarrow 0$ the physical mean value $\la \lambda \ra$ approaches its unperturbed magnitude $\lambda_0$ corresponding to the stationary Poisson distribution. However, in case when  $(t_0/T)\sim  1$ (which will be the case as we discuss below) the dimensionless parameter $(\la \lambda \ra/\lambda_0)$ must be smaller than one as it accounts for non-stationary Poisson processes. The parameter $(\la \lambda \ra/\lambda_0)\rightarrow 1$ approaches identity if  
non-stationary Poisson processes play the minor role. The deviation of this parameter from $(\la \lambda \ra/\lambda_0)\neq  1$ is a precise quantitative characteristic of the  non-stationary Poisson processes   in the dynamics of the system.  

From the basic features of the AQN model one should expect $(t_0/T)$ to be large, of order one. This is because a single AQN event could produce a number of radio emission events which should correspond to the  clustering events, since they are not independent. Furthermore, we also expect that $(\la \lambda \ra/\lambda_0)$ strongly deviates from the identity, which represents a quantitative  characteristic of a contribution due to the clustering events as the Poissonian distribution is characterized by a single parameter $\lambda_0$ with $\la \lambda \ra=\lambda_0$.

With this  preliminary remarks on physical meaning of the parameters we can now proceed with computations with the main goal to analyze the role of non-stationary Poisson processes in the radio wave emission as a result of the AQN annihilation events.  

One can combine equations (\ref{lambda1}) and (\ref{lambda2}) to represent $f(\lambda) $
as follows:
\be
\label{lambda3}
f(\lambda)&=&\left[\left(\frac{T-t_0}{T}\right)\delta(\lambda-\lambda_0)\right]\\
&+&\left[\frac{\beta+1}{\lambda_0}\frac{t_0}{T}\left(\frac{\lambda}{\lambda_0}\right)^{\beta}\theta(\lambda_0-\lambda)\right], \nonumber
\ee
where factor $({T-t_0})/{T}$ is inserted in front of delta function $\delta(\lambda-\lambda_0)$ to preserve the normalization (\ref{f}).

One should emphasize that the $\lambda_0$ is not the mean event occurrence rate $\la \lambda \ra$ anymore. Instead, the proper value for $\la \lambda \ra$ reads:
\be
\label{lambda4}
\la \lambda \ra\equiv \int \lambda f(\lambda)d\lambda = \lambda_0\left[1-\frac{1}{\beta+2}\left(\frac{t_0}{T}\right)\right].
\ee

Now we are in position to compute $ P (\Delta t)$ as defined by (\ref{P(t)3}):
\be
\label{P(t)4}
 P (\Delta t)=\frac{1}{\la \lambda \ra}{\int_0^{\infty}\lambda^2f(\lambda) e^{-[\lambda \Delta t]}d\lambda},
 \ee
 with $f(\lambda)$ as given by (\ref{lambda3}). The result can be represented as follows:
 \be
\label{P_final}
&& P (\Delta t)=\frac{\lambda_0^2 }{\la \lambda \ra} e^{-[\lambda_0 \Delta t]}\cdot\left(\frac{T-t_0}{T}\right) \\ &+&\frac{(\beta+1)\lambda_0^2}{\la \lambda \ra}  \cdot \left(\frac{t_0}{T}\right)\left[\int_0^{\lambda_0}  \frac{d\lambda }{\lambda_0}  \left(\frac{\lambda}{\lambda_0}\right)^{\beta+2} 
 e^{-[\lambda \Delta t]} \right], \nonumber 
  \ee
 where the first term describes the stationary Poisson distribution while the second term describes the deviation from 
 Poisson distribution  at small time scales. The second term in distribution (\ref{P_final}) can be expressed in terms of the lower incomplete $\gamma(s,x)$  function defined as follows:
 \be
 \label{gamma}
 \gamma (s, x)\equiv \int_0^xu^{s-1}e^{-u}du, ~~  \gamma (s, x)=\Gamma(s)-\Gamma(s,x),~~~
 \ee
 where $\Gamma(s)$ is the gamma function and $\Gamma(s,x)$ is the upper incomplete gamma function.
 We identify the parameters from the integrand entering (\ref{P_final}) as follows: 
 \be
 u=\lambda \Delta t, ~~~ x\equiv \lambda_0 \Delta t, ~~~ s=\beta+3
 \ee
  to arrive to the following expression for  $P (\Delta t)$ in terms of the lower incomplete $\gamma(s,x)$  function:
  \be
  \label{P_final_1}
  && P (\Delta t)=\frac{\lambda_0^2 }{\la \lambda \ra} e^{-[\lambda_0 \Delta t]} \cdot \left(\frac{T-t_0}{T}\right)\\ &+& 
  \frac{\lambda_0^2(\beta+1)}{\la \lambda \ra}  \cdot \left(\frac{t_0}{T}\right)\cdot \left(\frac{1}{\lambda_0\Delta t }\right)^{\beta+3}\cdot \gamma\left[\beta+3, \lambda_0 \Delta t\right]. \nonumber 
  \ee
  This expression is correct for any value of $t_0/T$.
  However, it is very instructive to see explicit dependence on $\Delta t$ when $t_0/T\ll 1 $ is small, and the Poisson distribution is restored.  
    
With this purpose in mind we simplify expression (\ref{P_final_1})by expanding  the incomplete gamma function entering (\ref{P_final_1}). Therefore,  the expression (\ref{P_final_1}) can be simplified as follows:
\be
  \label{P_final_2}
  && P (\Delta t)\approx\frac{\lambda_0^2 }{\la \lambda \ra} e^{-[\lambda_0 \Delta t]} \left(\frac{T-t_0}{T}\right)\\ &+& 
  \frac{\lambda_0^2(\beta+1)\Gamma(\beta+3)}{\la \lambda \ra}  \cdot \left(\frac{t_0}{T}\right)\cdot \left(\frac{1}{\lambda_0\Delta t }\right)^{\beta+3},  \nonumber 
  \ee
 where we use the identity (\ref{gamma}) and ignored the exponentially small contribution coming from incomplete upper gamma function:
 \be
 \Gamma (s, x\rightarrow \infty)\rightarrow x^{s-1} \exp (-x).
 \ee
 \exclude{
 The expression (\ref{P_final_2}) can be further simplified 
 \be
  \label{P_final_3}
   P (\Delta t)\approx\frac{\lambda_0 }{(1+\xi)} \left[e^{-[\lambda_0 \Delta t]}   +\xi\frac{ (\beta+2)\Gamma(\beta+3)}{(\lambda_0\Delta t )^{\beta+3}} \right],~~~
  \ee
  where we introduced a single parameter $\xi$ which describes the deviation from conventional Poisson distribution, and  defined as
  \be
  \label{xi}
  \xi\equiv \left(\frac{t_0}{T}\right)\cdot\left(\frac{\beta+1}{\beta+2} \right) \ll 1.
  \ee
  }
 In the limit $(t_0/T)\rightarrow 0$ we recover the conventional Poisson distribution, while $(t_0/T)\neq  0$ describes the deviation from 
 Poisson statistics in this simplified setting. 
 
 We are now ready to analyze the non-Poisson distribution given by (\ref{P_final_2}).
 Important  point here is that this distribution is a superposition of two parts: The first term describes the Poisson distribution
 with small correction in normalization. Most important part for us is the second term which is parametrically small at $(t_0/T)\ll 1$. However, it could become the dominant  part of the distribution $ P (\Delta t)$  at small $\Delta t\rightarrow 0$ due to a high power $(\Delta t)^{-(\beta+3)}$  in the denominator (\ref{P_final_2}).
 
 It is interesting to note that \cite{Mondal-2020} noticed that their data can be fitted as a superposition of two terms which have precisely the form of two terms entering (\ref{P_final_2}).
However, \cite{Mondal-2020} also fitted the observed signal to an expression which represents    the product of two terms rather than in form of sum of two terms entering  
the eq.  (\ref{P_final_2}) with well-defined physical meaning of the relevant parameters such as $(t_0/T)$. In next subsection we 
fit that data from \cite{Mondal-2020} using exact (\ref{P_final}) and simplified (\ref{P_final_2}) expressions for $P (\Delta t)$. Our main conclusion of this fit is that the clustering events play the dominant role in the distribution  $P (\Delta t)$.

\subsection{Wait time  distribution. Theory confronts the observations.}\label{confronts}
We are now in  position to compare the   occurrence probability  presented on Fig. 7 in \cite{Mondal-2020} with our theoretical formula 
(\ref{P_final}) which deviates from Poisson distribution as it includes the clustering events.

First, we have to comment that the occurrence probability plotted on 
Fig. 7 in \cite{Mondal-2020} is different from the wait time distribution $P(\Delta t)$  defined in the previous subsection. It is convenient to explain the difference using the description in terms of the discrete  bins $[\Delta t_i, \Delta t_{i+1}]$. In these terms, Fig. 7 of \cite{Mondal-2020} is a histogram, where the blue  points  represent the values $n_i/N$ where $n_i$ is the number of events with wait time located in the bin  $[\Delta t_i, \Delta t_{i+1}]$ and $N$ is the total number of events. However, by definition, the wait time distribution $P(\Delta t_i)$ is obtained by dividing $n_i/N$ by the bin width $[\Delta t_{i+1}-\Delta t_i]$ for proper normalization of $P(\Delta t_i)$. Indeed, 
\be \label{eq:Pi}
&&P(\Delta t_i)\equiv \frac{n_i}{N}\frac{1}{[\Delta t_{i+1}-\Delta t_i]},\\   &&\sum_i P(\Delta t_i) [\Delta t_{i+1}-\Delta t_i]=\sum_i \frac{n_i}{N}=1.\nonumber
\label{P_definition}
\ee

\begin{figure}
    \centering
    \includegraphics[width=1\linewidth]{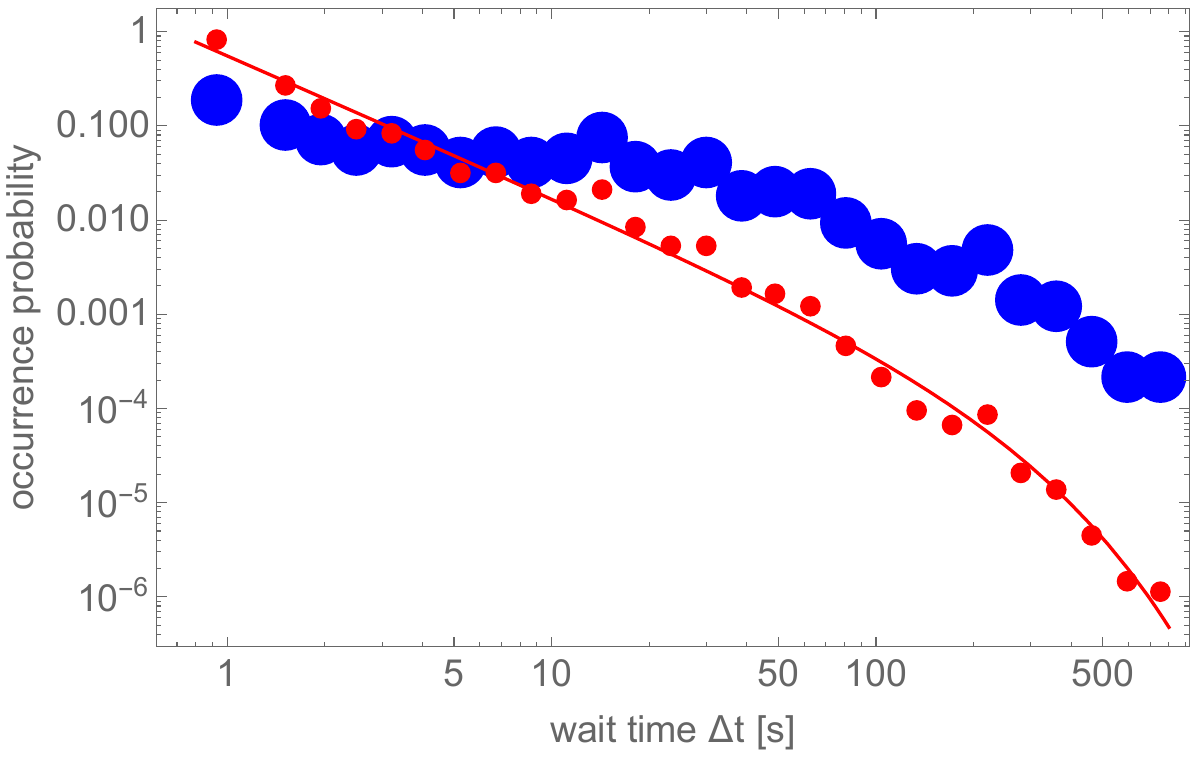}
    \caption{The blue points are extracted from Figure~7 in \cite{Mondal-2020} (132~MHz). Dividing the blue points by the corresponding bin width, we get the red points (i.e., the values of $P(\Delta t_{i})$ in (\ref{eq:Pi})). The red line is fitted by (\ref{eq:fit_Mondal}) with $A=0.56 s^{-1}, n\simeq 1.5, \lambda\simeq 0.0049 s^{-1}$.}
    \label{fig:Pi}
\end{figure}

As noticed by \cite{Mondal-2020}, the data can be nicely fitted using the following function 
\be
\label{fit}\label{eq:fit_Mondal}
P(\Delta t)=A (\Delta t)^{-n}\exp(-\lambda \Delta t),
\ee
where the continuum limit is already assumed.
We confirm that the good match can indeed be achieved, and the corresponding fit is shown by the red line on Fig.~\ref{fig:Pi}.

\begin{figure}
    \centering
    \includegraphics[width=1\linewidth]{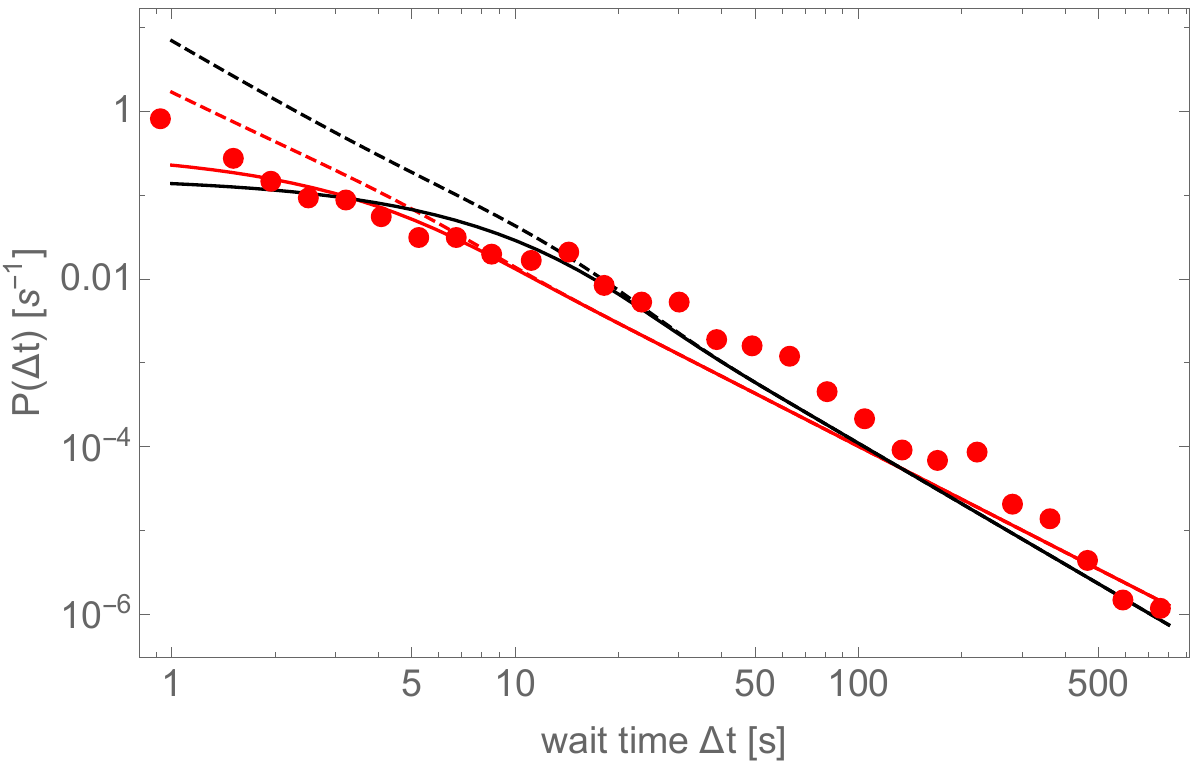}
    \caption{The red points are the same as those in Fig.~\ref{fig:Pi} (i.e., the values of $P(\Delta t_{i})$ in (\ref{eq:Pi})). The solid line are fitted by the full expression of $P(\Delta t)$ given  by  (\ref{P_final}). The solid red line gives $\beta = -0.9$, $t_0 = 4000$~s, $\lambda_0 = 0.5{\rm~s}^{-1}$. Other choices around this group of parameters can also give similar result. For example, the solid black line corresponds to $\beta = -0.6$, $t_0 = 3000$~s, $\lambda_0 = 0.2{\rm~s}^{-1}$. In comparison, the dashed lines are the simplified $P(\Delta t)$ given by (\ref{P_final_2}), with the same group of parameters chosen correspondingly.}
    \label{fig:Pt}
\end{figure}

 We are now ready to interpret the results obtained above in terms of the two dimensionless  parameters $(t_0/T) $  and $(\la \lambda \ra/\lambda_0)$
introduced in (\ref{lambda2}) as the generic way to parameterize  non-stationary Poisson processes. First of all, the acceptable  fit shown on Fig  \ref{fig:Pt}
always produces the relatively large value for $(t_0/T)$. Indeed, the first solution corresponds to $(t_0/T)\simeq 0.95 $, while the second solution assumes the value 
$(t_0/T)\simeq 0.71 $. We remind that this parameter  $(t_0/T)$ describes the portion of time when clustering events occur. In case of stationary Poisson processes, $(t_0/T)=0$. Fit in both cases suggests that non-stationary Poisson processes occur for most of the time,  which unambiguously implies that non-stationary Poisson processes play the dominant role in radio wave emission. This is consistent with the AQN proposed mechanism when the non-stationary Poisson distribution is expected and anticipated. 

Another quantitative characteristic which describes the deviation from conventional Poisson distribution is the dimensionless  parameter $(\la \lambda \ra/\lambda_0)$. 
The acceptable  fit shown on Fig  \ref{fig:Pt}
always produces a strong deviation of the parameter $(\la \lambda \ra/\lambda_0)$ from identity. Indeed, $(\la \lambda \ra/\lambda_0)\approx 0.14$ for the first solution while $(\la \lambda \ra/\lambda_0)\approx 0.5$ for the second solution. This represents another  strong evidence supporting our claim that non-stationary Poisson processes play the dominant role in radio wave emission. 

One should note that the quantitative estimates of the parameters from the first principles  within the AQN framework are hard to carry out. We could only anticipate a large deviation from conventional Poisson distribution as explained above, while any quantitative estimates of these parameters are not feasible at the moment. The problem  is that the radio emission  by non-thermal electrons is a random process, which strongly depends on surrounding plasma features. Furthermore, the emission spectrum of the non-thermal electrons also represents a challenging theoretical problem, as emission occurs in the system which is moving with very large Mach number when the turbulence, shock waves and other non- equilibrium processes dominate the dynamics of the non-thermal electron's emission.

Our first quantitative prediction is that the parameters $(t_0/T) $ and $(\la \lambda \ra/\lambda_0)$ must be very similar for different frequency bands. 
Our second quantitative prediction, is that  the emission between radio events observed at  different frequencies  must be correlated with time delays measured in seconds. This correlation is very specific to the AQN mechanism (see item 4 in Conclusion).
 
It is interesting to note 
  that  the data from \cite{Mondal-2020}   can be nicely fitted using the   function  (\ref{fit}), which exhibits structures similar to our formula (\ref{P_final_2}).
  The important  difference here is that our formula was derived with well defined parameters   $(t_0/T) $  and $(\la \lambda \ra/\lambda_0)$ , which   quantitatively characterize  the   non-stationary Poisson processes, while the extraction of $A, \lambda, n$ from the fitting  (\ref{fit}) represented on Fig. \ref{fig:Pi} does not allow to arrive to any quantitative conclusion.

As clustering events play a major role, one may wonder if our estimate of $\bar{B}$ in section \ref{rate} may be modified as a result of these events. We think that the corresponding variation is numerically mild, and does not modify the picture advocated in this work\footnote{Indeed, even if each AQN event generates a cluster consisting on average, let us say,   three radio events, it would change the event rate (\ref{eq:observation_rate}) by the same factor three. We note, that much larger number of events within the same cluster would be inconsistent with total energy estimate (\ref{flux-radio}) which agrees with observations (\ref{flux-total}).  The scaling parameter $\alpha\simeq 2.5$ defined by (\ref{eq:groups}) implies that the corresponding variation in $\bar{B}$ does not exceed a factor $\sqrt[2.5]{3}\approx 1.5$. These changes are much smaller than the difference in $\bar{B}$ between distinct  acceptable models (\ref{eq:groups}) , as one can see from Fig. \ref{fig:impact_rate}.}. Therefore, we ignore the corresponding modifications in $\bar{B}$ in the present study.

We have discussed at length that the presence of clustering  events  is a generic feature   of the mechanism for  impulsive radio events. We interpret the fit shown on Fig.~\ref{fig:Pt}  of the data from \cite{Mondal-2020} with our expression (\ref{P_final}) as an additional strong support for our proposal when  radio emissions always accompany nanoflare events.

One should emphasize that nanoflares are introduced as generic   events, producing an impulsive energy release at small scale ( see the review papers \cite{Klimchuk:2005nx,Klimchuk:2017}). \textit{The fact that nanoflares are the consequence of AQN annihilation events  accompanied by the clustering of radio events is a highly nontrivial consistency check of the entire framework .} Such clustering events supported by data \cite{Mondal-2020} are clearly related to a non-Poissonian character of distribution, and the AQN model provides a natural solution for this feature.

\section{Conclusion and Future development}\label{conclusion}
    We proposed that AQN annihilation events can be identified with nanoflares and showed that they are inevitably accompanied by radio events. This  proposal is consistent 
    with all observations reported by \cite{Mondal-2020}, including  the  frequency of appearance, the temporal and spatial distributions, \textit{their intensity}, and other related observables.  
There are several direct consequences of this proposal, which future observations will be able to support or refute:\\
1. The proposed mechanism  suggests that a considerable portion of radio events, recorded at different frequencies, might be emitted by a single AQN continuously generating  radio signals, as a result of different plasma frequencies at different altitudes.  This picture suggests that there must be a spatial  correlation between radio events in a given local patch (with size  $\sim 10^5 \rm km$ ), in the  different frequency bands, with time delays measured in seconds. 

Observations of correlated clustering events, as discussed in subsections \ref{clustering}, \ref{confronts} , are the direct manifestation of correlations observed in the same frequency band. We advocate the idea that 
similar spatial correlations, from different frequency bands, must  also exist. This prediction can be directly tested by  MWA.  There must also be similar temporal correlations (see item 4 below).

2. Lower frequency waves could be emitted from higher altitudes. The important point here being that the dependence of the intensity of the emission on  the  altitude  is a highly nontrivial  function, for several reasons.    
 First, the upward moving non-thermal electrons are much more numerous
at lower altitude (corresponding to  higher  $\nu$) in comparison to higher altitudes (corresponding to lower $\nu$) because the $n_s/n_e$ ratio scales as $r^{-3/2}$. When this scaling reaches a ratio below the required rate (\ref{n_s2}) , the radio wave emission cannot occur, as the density of the non-thermal electrons is not sufficient for the plasma instability to develop \cite{Thejappa-1991}. Furthermore, the effective mean free path  determined by  (\ref{lambda_eff}) defines the highest altitudes the non-thermal electrons may  reach. Above this altitude, the non-thermal electrons will thermalize  and cannot be a source of the radio waves.  

As a result of these suppression factors,   we expect that   the low frequency emissions should be, in general,    suppressed. Of course, the radio emission is related to   random processes, and highly sensitive to some specific local features of the plasma and non-thermal electrons, 
as discussed  in subsection \ref{mechanism}.  Therefore, our prediction on suppression is the subject of possible fluctuations   within small frequency bands.
This tendency has been  indeed observed for the 98 MHz band,   where the recorded number of resolved events is at least one order of magnitude smaller than for the three other higher frequencies  bands.  We predict that the emission rate  at 80 MHz and 89 MHz, which have been recorded, but are not yet published by \cite{Mondal-2020}, should  demonstrate  even lower rate of resolved events (even in comparison with 98 MHz emission). This is a highly nontrivial prediction  of our proposal,  as it is difficult to understand this behavior using alternative models, since the electron density in the corona is  a very smooth function in this region (see Fig. \ref{fig:frequency}). This prediction can be directly tested by  MWA, as the observations according to 
\cite{Mondal-2020} were done in 12 frequency bands, including the low frequencies of the 80 and 89 MHz bands.  

3. In contrast with the low frequency bands, the event rate for higher frequency bands should be higher than the rate recorded for the 160 MHz band .   This prediction  can be directly tested in  future analysis  by studying emissions with $\nu\gtrsim 160$ MHz  since, according to 
\cite{Mondal-2020} , some of their observations were done in the 179, 196, 217 and 240 MHz bands.  One should comment here that at higher frequencies ( $\nu\gtrsim$ 240 MHz) , radio emission is subject to a strong absorption, and that the  observed intensity  will experience suppression \cite{oberoi_2017},  limiting our perspectives  to study higher frequency emissions.

4.  The  proposed AQN mechanism of radio emission  predicts the presence of correlations between the emissions at different frequency bands. These correlations emerge due to the upward motion of the non-thermal electrons, with typical velocities $v_{\perp}\simeq \rm 10^4 km/s$ according to (\ref{N_e3}). 
The delays in arrival time at different heights is measured in seconds, when heights vary on the scale of  $10^4$ km, according to Fig. \ref{fig:frequency}. As a result of upward motion, the low frequency emissions should be delayed in comparison to the  high frequency emissions. Observing these correlated radio emissions in different frequency bands would 
unambiguously support our proposal, as it is very hard to imagine how such correlations could occur in any alternative scenarios.  

5. Solar Orbiter recently observed so-called ``campfires"   in the  extreme UV frequency bands. It is tempting to identify such events with  the annihilation of large sized  AQNs, as they are capable of generating radio signals sufficiently strong to be resolved. We therefore suggest to  search for a cross correlation between MWA radio signals and recordings of the extreme UV photons by Solar Orbiter.

 \section*{Acknowledgements}

This research was supported in part by the Natural Sciences and Engineering
Research Council of Canada. 

\appendix
\section{Simulations} \label{appendix:simulations}
The appendix shows the details of the MC simulation implemented in this work.  

\subsection{The simulation setup}
The setup of the simulation in the present work follows that in \cite{Raza:2018gpb}, which can be divided into three steps. The first step is to use the MC method to generate a large number of dark matter particles in the solar neighborhood and collect the ones that will eventually impact the Sun. The second step is to assign AQNs masses to the particles. We will use different models of the AQN mass distributions (as shown in (\ref{eq:groups})). The third step is to solve the multiple differential equations that dominate the annihilation process of AQNs in the solar atmosphere.

\textit{Step 1}. In this step, we simulate the positions and velocities of dark matter particles in the solar neighborhood. The velocity distribution of the dark matter particles, with respect to the solar system frame, follows a Maxwellian distribution:
\begin{equation}\label{eq:maxwellian}
    f_{\vec{v}}(\vec{v}) d^{3} \vec{v} = \frac{d^{3} \vec{v}}{(2\pi\sigma^2)^{3/2}}\exp\left[-\frac{v_{x}^2+v_{y}^2+(v_{z}-v_{\odot})^2}{2\sigma^2}\right]
\end{equation}
where the velocity dispersion is $\sigma\simeq110~{\rm km}/{\rm s}$, and the velocity shift $v_{\odot}\simeq 220~{\rm km}/{\rm s}$ is due to the relative motion between the Sun and the dark matter halo.

\begin{figure*}
    \centering
    \includegraphics[width=0.8\linewidth]{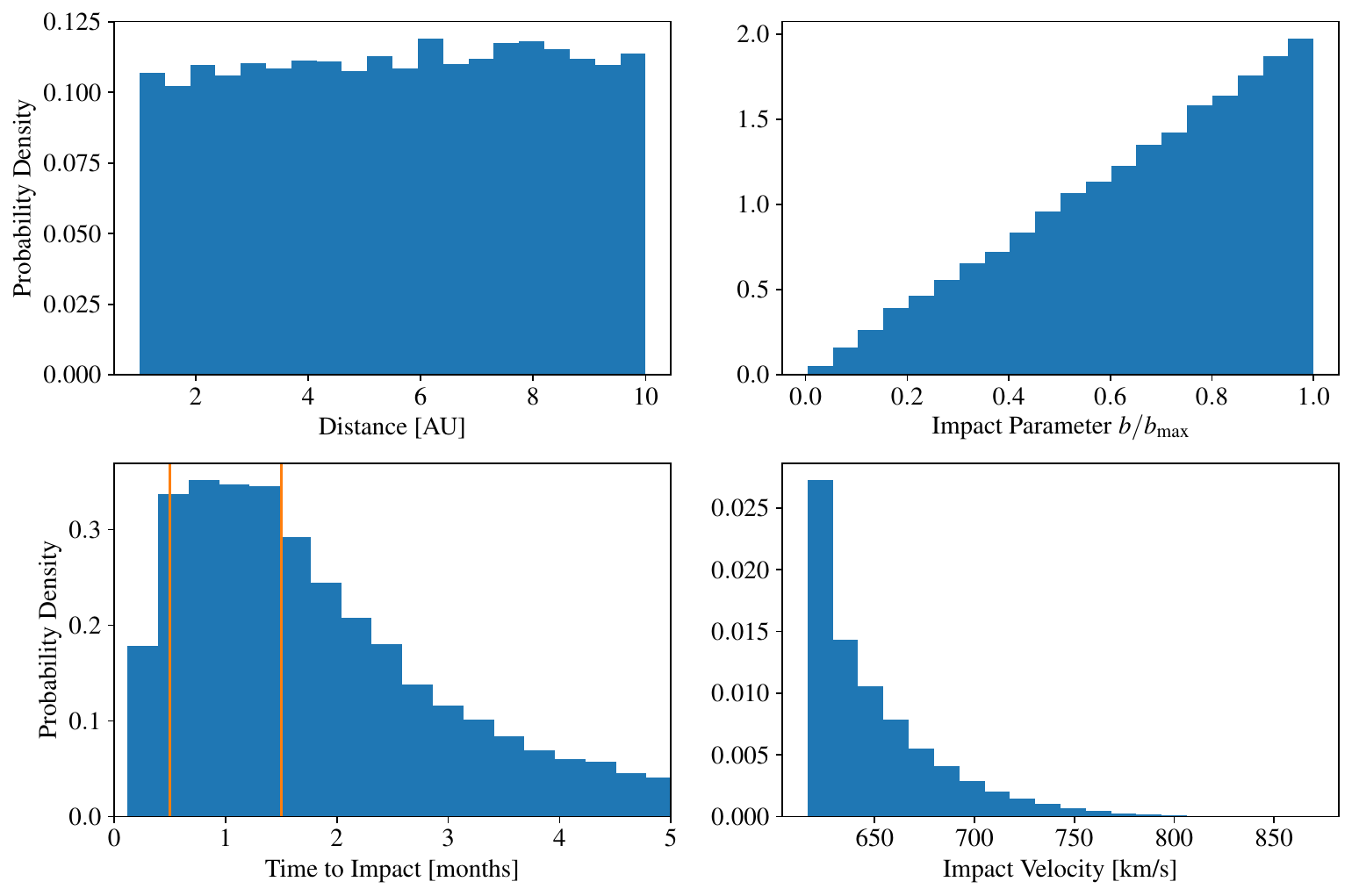}
    \caption{Probability density distributions of the trajectory and impact properties for the $N_{\rm imp}=30457$ impacting particles. Starting from the top-left panel to the bottom-right, the plots represent (a): the initial distance distribution of these impacting particles. (b): the impact parameter distribution. (c): the velocity distribution when they impact the Sun. (d): the impact time distribution.}
    \label{fig:distribution}
\end{figure*}

The positions of particles are generated in such a way that they initially \textit{uniformly} populate in a spherical shell around the Sun. The inner and outer boundaries of the spherical shell are respectively $R_{\rm min}=1$ AU and $R_{\rm max}=10$ AU. Note that our choice of $R_{\rm min}$ is different from Ref.~\cite{Raza:2018gpb} where $R_{\rm min}=R_{\odot}$ there. Choosing a larger $R_{\rm min}$ is to reduce the effect of the Sun's gravity on the \textit{initial} velocity distribution~(\ref{eq:maxwellian}). The solar escape velocity at $1$~AU is $v_{e}\approx 42~{\rm km}/{\rm s}$, so when a particle moves from infinity with the typical velocity $v_{0}=220~{\rm km}/{\rm s}$ to this distance, the velocity increment due to the Sun's gravity is $\Delta v = \sqrt{v_{0}^2+v_{e}^2}-v_{0}\approx 4~{\rm km}/{\rm s}$ which is very small compared with $v_0$.
Similar to Ref.~\cite{Raza:2018gpb}, we generated $N_{\rm sample}= 2\times 10^{10}$ such particles. The particles then move following Newton's gravity, attracted by the Sun, using the classical two-body orbit dynamics. The criteria to determine whether or not a particle will impact the Sun are also the same as in \cite{Raza:2018gpb}. For a given particle, if the perihelion of the hyperbolic trajectory is smaller than $R_{\odot}$ (and also if the velocity direction is inward), then it will impact the Sun. It turns out that from the initial sample of $2\times10^{10}$, the number of particles that will impact the Sun is $N_{\rm imp}=30457$.\footnote{In comparison, the number obtained  in Ref.~\cite{Raza:2018gpb} is 36123. The difference is beyond the statistical fluctuation. This difference occurs not only due to our choice of a larger $R_{\rm min}$, the inner boundary of the initially simulated region, but also a technical detail that a different method (more appropriate) is chosen in determining the perihelion. However, all of these changes have no significant effects on the results as we can see in Fig.~\ref{fig:distribution} by comparing it with Ref.~\cite{Raza:2018gpb}.} The trajectory and impact properties of these impacting particles are shown in Fig.~\ref{fig:distribution}.

The expression of the impact parameter $b$ is
\begin{equation}
    b = r_{p} \sqrt{1+\frac{2GM_{\odot}}{r_{p} v_{0}^2}}
\end{equation}
where $r_{p}$ is the perihelion distance. $v_0$ is the particle velocity at infinity that can be extrapolated from the initial velocity and position simulated, i.e. $v_0=\sqrt{v_i^2 - 2GM_{\odot}/r_{i}}$. The impacting requires $0\leq r_p \leq R_{\odot}$. If we take $r_{p}=R_{\odot}$, we get the maximum impact parameter $b_{\rm max}$. The distribution of the impact parameter (in the form of $b/b_{\rm max}$) is shown in the subplot (b) of Fig.~\ref{fig:distribution}.

From the distribution of impact time as shown in the subplot (c) of Fig.~\ref{fig:distribution}, we can calculate the impact rate. Following the logic in Ref.~\cite{Raza:2018gpb}, we should only use the time window where the rate is constant. We choose it as $t_{\rm imp}\in [0.5, 1.5]$~months where the boundaries are denoted as two vertical lines in the plot. The rate in the time window is constant because the dominant part of particles impacting the Sun are the particles from the the spherical shell between $R_{\rm min}$ and $R_{\rm max}$. Outside the time window, we see the rate drops because we did not simulate the particles outside the spherical shell. The impact rate is $N(\Delta t_{\rm imp})/\Delta t_{\rm imp}$ where $N(\Delta t_{\rm imp})$ is the number of particles impacting the Sun in the time window chosen above. Note that this impact rate is not the true impact rate because the number of AQNs simulated, $N_{\rm sample}= 2\times 10^{10}$, is not the true number of AQNs inside the spherical shell. 

To convert the impact rate to the true impact rate, we need to multiply it by the scaling factor $f_{S}$ which is the ratio of the true number of AQNs in the spherical shell to $N_{\rm sample}$:
\begin{equation}\label{eq:scaling_factor}
    f_{S} = \frac{\frac{4}{3}\pi(R_{\rm max}^2-R_{\rm min}^2)\cdot n_{\rm AQN}}{N_{\rm sample}}
\end{equation}
where $n_{\rm AQN}$ is the number density of antimatter AQNs in the solar system:
\begin{equation}
   n_{\rm AQN} = \left(\frac{2}{3}\cdot\frac{3}{5}\cdot \rho_{\rm DM}\right) \cdot \frac{1}{m_{P}\left<B\right>}. 
\end{equation}
and $\rho_{\rm DM}\simeq 0.3 {\rm~GeV~cm^{-3}}$ is the dark matter density in the solar system. $3/5$ of the dark matter is in the form of antimatter AQN; $2/3$ mass of an AQN is in the form of baryons (the remaining $1/3$ is in the form of axions). $m_P$ is the proton mass. $\left<B\right>$ is the average baryon number carried by an AQN. It depends on different models of AQN mass distribution that will be discussed in Step 2. Thus, the true rate of (antimatter) AQNs impacting the sun is
\begin{equation}\label{eq:impact_rate}
    \frac{dN}{dt} = \frac{N(\Delta t_{\rm imp})}{\Delta t_{\rm imp}}\cdot f_{S}, ~~~~t_{\rm imp}\in [0.5, 1.5] {\rm ~months}.
\end{equation}

\textit{Step 2}. We are now assigning masses (baryon numbers) to all the AQNs collected in Step 1 that will impact the Sun. For each AQN, its mass is assigned randomly with the probability following one of the three models of power-law distribution, (\ref{eq:groups}). Thus, we have three copies of all the impacting AQNs with different mass distributions. 

\textit{Step 3}. The evolution of an AQN in the solar atmosphere is described by a system of differential equations, including the kinetic energy loss due to friction and the mass loss due to the annihilation events of the antibaryons carried by AQN with the baryons in the atmosphere. We refer the reader to \cite{Raza:2018gpb} for a complete list of the differential equations needed here, and their derivation. In order to solve these equations numerically, the density and temperature profiles of the solar atmosphere above the photosphere is also needed. In this work, we adopt the profiles presented in Ref.~\cite{2008ApJS..175..229A} which are more accurate than those used in \cite{Raza:2018gpb}.

The mass loss varying with time (or equivalently, height above the solar photosphere) for the $N_{\rm imp}$ AQNs is then computed numerically.

\subsection{Results}
The results obtained from the numerical simulations are presented in the main text. Additional details are given here.

Fig.~\ref{fig:impact_rate} shows the rate of AQNs in the mass range $[\bar{B}, B_{\rm max}]$ impacting the Sun. The rate is calculated as (\ref{eq:impact_rate}) but with only large AQNs $B\geq \bar{B}$ taken into account. By varying the value of the cutoff $\bar{B}$ from $B_{\rm min}$ to $B_{\rm max}$, we quantify how the impact rate depends on $\bar{B}$, as shown in the figure. The rate at $\bar{B}=B_{\rm min}$ is the total impact rate. For the three groups, the total impact rates are respectively $4.17\times 10^4 {\rm ~s^{-1}}$, $1.52\times 10^4 {\rm ~s^{-1}}$ and $3.52\times 10^3 {\rm ~s^{-1}}$ which match well Ref.~\cite{Raza:2018gpb} (see Figure 8 there).


In addition, the luminosity $L^{\odot}$ can be calculated as $L^{\odot}= 2\left<\Delta m\right>c^2 \cdot dN/dt$ where $\Delta m$ is an AQN's mass loss along its trajectory through the solar atmosphere before entering the dense region, the photosphere. Similarly, we can compute the luminosity $L^{\odot}_{\bar{B}}$ by counting large AQNs ($B\geq \bar{B}$) only, and the result is shown in Fig.~\ref{fig:impact_luminosity} for different groups of mass distribution. The total luminosity is obtained at $\bar{B}=B_{\rm min}$. For the three groups, the total luminosity are respectively $1.05\times10^{27} {\rm ~erg\cdot s^{-1}}$, $1.07\times10^{27} {\rm ~erg\cdot s^{-1}}$ and $1.06\times10^{27} {\rm ~erg\cdot s^{-1}}$ which match well Ref.~\cite{Raza:2018gpb} (see Figure 10 there). 

One may notice that in the two left subfigure of Fig.~\ref{fig:impact_rate} and Fig.~\ref{fig:impact_luminosity}, the simulated lines become zigzag at large baryon numbers. This is because the proportion of large AQNs is actually very small. Despite the number of all the impacting AQNs is as large as $30457$, the power-law index $\alpha \sim (2-2.5)$ makes the hit rate with large $B$ very tiny when assigning masses to AQNs randomly in Step 2. For example, our statistical result shows that in Group 1, the number of AQNs with $B\geq 5\times 10^{26}$ is only 12, and the the number of AQNs with $B\geq 10^{27}$ is only 3. Such tiny number causes large statistical fluctuation, so we see the zigzags in the two left subfigures. We have to generate enough large AQNs to remove the large statistical fluctuation. 

We resolve this technical problems as follow. We  simulate \textit{another} $10^{10}$ AQNs by redoing the three steps in the setup as described above. We call this procedure as the second-round simulation. We get $15019$ AQNs that will finally impact the Sun out of the total $10^{10}$ AQNs. The masses (baryon numbers) assigned to these $15019$ impacting AQNs are constrained in the range $B\in[B_{L}, B_{\rm max}]$. $B_{L}$ for each group should be chosen well above $B_{\rm min}$ to ensure that enough large AQNs can be generated, but $B_{L}$ should not exceed the start of the zigzags. Although we did not simulate all AQNs in this second-round simulation, we can extrapolate the ``number'' of impacting AQNs in the \textit{full} mass range by looking at the proportion of large AQNs ($B\in[B_{L}, B_{\rm max}]$) in the full mass range~\footnote{The advantage of assigning AQNs masses only in the range of $[B_{L}, B_{\rm max}]$ is that we don't repeat generating a huge amount of small AQNs which are far more than needed and only to make the simulations extremely time-consuming.}. Furthermore, we can calculate the extrapolated $N_{\rm sample}$ and the extrapolated scaling factor $f_{S}$. Finally, we obtain the true impact rate of these large AQNs simulated in the second-round simulation. Similarly, we obtain the luminosity. The results are shown in the two right subfigures of Fig.~\ref{fig:impact_rate} and Fig.~\ref{fig:impact_luminosity}, where we see that the large statistical fluctuation disappears.

\bibliography{radio_corona}

\begin{thebibliography}{71}%
\makeatletter
\providecommand \@ifxundefined [1]{%
 \@ifx{#1\undefined}
}%
\providecommand \@ifnum [1]{%
 \ifnum #1\expandafter \@firstoftwo
 \else \expandafter \@secondoftwo
 \fi
}%
\providecommand \@ifx [1]{%
 \ifx #1\expandafter \@firstoftwo
 \else \expandafter \@secondoftwo
 \fi
}%
\providecommand \natexlab [1]{#1}%
\providecommand \enquote  [1]{``#1''}%
\providecommand \bibnamefont  [1]{#1}%
\providecommand \bibfnamefont [1]{#1}%
\providecommand \citenamefont [1]{#1}%
\providecommand \href@noop [0]{\@secondoftwo}%
\providecommand \href [0]{\begingroup \@sanitize@url \@href}%
\providecommand \@href[1]{\@@startlink{#1}\@@href}%
\providecommand \@@href[1]{\endgroup#1\@@endlink}%
\providecommand \@sanitize@url [0]{\catcode `\\12\catcode `\$12\catcode
  `\&12\catcode `\#12\catcode `\^12\catcode `\_12\catcode `\%12\relax}%
\providecommand \@@startlink[1]{}%
\providecommand \@@endlink[0]{}%
\providecommand \url  [0]{\begingroup\@sanitize@url \@url }%
\providecommand \@url [1]{\endgroup\@href {#1}{\urlprefix }}%
\providecommand \urlprefix  [0]{URL }%
\providecommand \Eprint [0]{\href }%
\providecommand \doibase [0]{http://dx.doi.org/}%
\providecommand \selectlanguage [0]{\@gobble}%
\providecommand \bibinfo  [0]{\@secondoftwo}%
\providecommand \bibfield  [0]{\@secondoftwo}%
\providecommand \translation [1]{[#1]}%
\providecommand \BibitemOpen [0]{}%
\providecommand \bibitemStop [0]{}%
\providecommand \bibitemNoStop [0]{.\EOS\space}%
\providecommand \EOS [0]{\spacefactor3000\relax}%
\providecommand \BibitemShut  [1]{\csname bibitem#1\endcsname}%
\let\auto@bib@innerbib\@empty
\bibitem [{\citenamefont {Mondal}\ \emph {et~al.}(2020)\citenamefont {Mondal},
  \citenamefont {Oberoi},\ and\ \citenamefont {Mohan}}]{Mondal-2020}%
  \BibitemOpen
  \bibfield  {author} {\bibinfo {author} {\bibfnamefont {S.}~\bibnamefont
  {Mondal}}, \bibinfo {author} {\bibfnamefont {D.}~\bibnamefont {Oberoi}}, \
  and\ \bibinfo {author} {\bibfnamefont {A.}~\bibnamefont {Mohan}},\ }\href
  {\doibase 10.3847/2041-8213/ab8817} {\bibfield  {journal} {\bibinfo
  {journal} {The Astrophysical Journal}\ }\textbf {\bibinfo {volume} {895}},\
  \bibinfo {pages} {L39} (\bibinfo {year} {2020})}\BibitemShut {NoStop}%
\bibitem [{\citenamefont {{Zhitnitsky}}(2017)}]{Zhitnitsky:2017rop}%
  \BibitemOpen
  \bibfield  {author} {\bibinfo {author} {\bibfnamefont {A.}~\bibnamefont
  {{Zhitnitsky}}},\ }\href {\doibase 10.1088/1475-7516/2017/10/050} {\bibfield
  {journal} {\bibinfo  {journal} {\jcap}\ }\textbf {\bibinfo {volume} {10}},\
  \bibinfo {eid} {050} (\bibinfo {year} {2017})},\ \Eprint
  {http://arxiv.org/abs/1707.03400} {arXiv:1707.03400 [astro-ph.SR]}
  \BibitemShut {NoStop}%
\bibitem [{\citenamefont {Raza}\ \emph {et~al.}(2018)\citenamefont {Raza},
  \citenamefont {van Waerbeke},\ and\ \citenamefont
  {Zhitnitsky}}]{Raza:2018gpb}%
  \BibitemOpen
  \bibfield  {author} {\bibinfo {author} {\bibfnamefont {N.}~\bibnamefont
  {Raza}}, \bibinfo {author} {\bibfnamefont {L.}~\bibnamefont {van Waerbeke}},
  \ and\ \bibinfo {author} {\bibfnamefont {A.}~\bibnamefont {Zhitnitsky}},\
  }\href {\doibase 10.1103/PhysRevD.98.103527} {\bibfield  {journal} {\bibinfo
  {journal} {Phys. Rev. D}\ }\textbf {\bibinfo {volume} {98}},\ \bibinfo
  {pages} {103527} (\bibinfo {year} {2018})},\ \Eprint
  {http://arxiv.org/abs/1805.01897} {arXiv:1805.01897 [astro-ph.SR]}
  \BibitemShut {NoStop}%
\bibitem [{\citenamefont {{Parker}}(1983)}]{Parker-1983}%
  \BibitemOpen
  \bibfield  {author} {\bibinfo {author} {\bibfnamefont {E.~N.}\ \bibnamefont
  {{Parker}}},\ }\href {\doibase 10.1086/160637} {\bibfield  {journal}
  {\bibinfo  {journal} {\apj}\ }\textbf {\bibinfo {volume} {264}},\ \bibinfo
  {pages} {642} (\bibinfo {year} {1983})}\BibitemShut {NoStop}%
\bibitem [{\citenamefont {{Zhitnitsky}}(2003)}]{Zhitnitsky:2002qa}%
  \BibitemOpen
  \bibfield  {author} {\bibinfo {author} {\bibfnamefont {A.~R.}\ \bibnamefont
  {{Zhitnitsky}}},\ }\href {\doibase 10.1088/1475-7516/2003/10/010} {\bibfield
  {journal} {\bibinfo  {journal} {\jcap}\ }\textbf {\bibinfo {volume} {10}},\
  \bibinfo {eid} {010} (\bibinfo {year} {2003})},\ \Eprint
  {http://arxiv.org/abs/hep-ph/0202161} {hep-ph/0202161} \BibitemShut {NoStop}%
\bibitem [{\citenamefont {{Grotrian}}(1939)}]{Grotrian-1939}%
  \BibitemOpen
  \bibfield  {author} {\bibinfo {author} {\bibfnamefont {W.}~\bibnamefont
  {{Grotrian}}},\ }\href {\doibase 10.1007/BF01488890} {\bibfield  {journal}
  {\bibinfo  {journal} {Naturwissenschaften}\ }\textbf {\bibinfo {volume}
  {27}},\ \bibinfo {pages} {214} (\bibinfo {year} {1939})}\BibitemShut
  {NoStop}%
\bibitem [{\citenamefont {{De Moortel}}\ and\ \citenamefont
  {{Browning}}(2015)}]{De-Moortel-2015}%
  \BibitemOpen
  \bibfield  {author} {\bibinfo {author} {\bibfnamefont {I.}~\bibnamefont {{De
  Moortel}}}\ and\ \bibinfo {author} {\bibfnamefont {P.}~\bibnamefont
  {{Browning}}},\ }\href {\doibase 10.1098/rsta.2014.0269} {\bibfield
  {journal} {\bibinfo  {journal} {Philosophical Transactions of the Royal
  Society of London Series A}\ }\textbf {\bibinfo {volume} {373}},\ \bibinfo
  {pages} {20140269} (\bibinfo {year} {2015})},\ \Eprint
  {http://arxiv.org/abs/1510.00977} {arXiv:1510.00977 [astro-ph.SR]}
  \BibitemShut {NoStop}%
\bibitem [{\citenamefont {{Klimchuk}}(2006)}]{Klimchuk:2005nx}%
  \BibitemOpen
  \bibfield  {author} {\bibinfo {author} {\bibfnamefont {J.~A.}\ \bibnamefont
  {{Klimchuk}}},\ }\href {\doibase 10.1007/s11207-006-0055-z} {\bibfield
  {journal} {\bibinfo  {journal} {\solphys}\ }\textbf {\bibinfo {volume}
  {234}},\ \bibinfo {pages} {41} (\bibinfo {year} {2006})},\ \Eprint
  {http://arxiv.org/abs/astro-ph/0511841} {astro-ph/0511841} \BibitemShut
  {NoStop}%
\bibitem [{\citenamefont {{Klimchuk}}(2017)}]{Klimchuk:2017}%
  \BibitemOpen
  \bibfield  {author} {\bibinfo {author} {\bibfnamefont {J.~A.}\ \bibnamefont
  {{Klimchuk}}},\ }\href@noop {} {\bibfield  {journal} {\bibinfo  {journal}
  {ArXiv e-prints}\ } (\bibinfo {year} {2017})},\ \Eprint
  {http://arxiv.org/abs/1709.07320} {arXiv:1709.07320 [astro-ph.SR]}
  \BibitemShut {NoStop}%
\bibitem [{\citenamefont {{Witten}}(1984)}]{Witten:1984rs}%
  \BibitemOpen
  \bibfield  {author} {\bibinfo {author} {\bibfnamefont {E.}~\bibnamefont
  {{Witten}}},\ }\href {\doibase 10.1103/PhysRevD.30.272} {\bibfield  {journal}
  {\bibinfo  {journal} {\prd}\ }\textbf {\bibinfo {volume} {30}},\ \bibinfo
  {pages} {272} (\bibinfo {year} {1984})}\BibitemShut {NoStop}%
\bibitem [{\citenamefont {{Madsen}}(1999)}]{Madsen:1998uh}%
  \BibitemOpen
  \bibfield  {author} {\bibinfo {author} {\bibfnamefont {J.}~\bibnamefont
  {{Madsen}}},\ }in\ \href {\doibase 10.1007/BFb0107314} {\emph {\bibinfo
  {booktitle} {Hadrons in Dense Matter and Hadrosynthesis}}},\ \bibinfo
  {series} {Lecture Notes in Physics, Berlin Springer Verlag}, Vol.\ \bibinfo
  {volume} {516},\ \bibinfo {editor} {edited by\ \bibinfo {editor}
  {\bibfnamefont {J.}~\bibnamefont {{Cleymans}}}, \bibinfo {editor}
  {\bibfnamefont {H.~B.}\ \bibnamefont {{Geyer}}}, \ and\ \bibinfo {editor}
  {\bibfnamefont {F.~G.}\ \bibnamefont {{Scholtz}}}}\ (\bibinfo {year} {1999})\
  p.\ \bibinfo {pages} {162},\ \Eprint {http://arxiv.org/abs/astro-ph/9809032}
  {astro-ph/9809032} \BibitemShut {NoStop}%
\bibitem [{\citenamefont {{Peccei}}\ and\ \citenamefont
  {{Quinn}}(1977)}]{axion1}%
  \BibitemOpen
  \bibfield  {author} {\bibinfo {author} {\bibfnamefont {R.~D.}\ \bibnamefont
  {{Peccei}}}\ and\ \bibinfo {author} {\bibfnamefont {H.~R.}\ \bibnamefont
  {{Quinn}}},\ }\href {\doibase 10.1103/PhysRevD.16.1791} {\bibfield  {journal}
  {\bibinfo  {journal} {\prd}\ }\textbf {\bibinfo {volume} {16}},\ \bibinfo
  {pages} {1791} (\bibinfo {year} {1977})}\BibitemShut {NoStop}%
\bibitem [{\citenamefont {{Weinberg}}(1978)}]{axion2}%
  \BibitemOpen
  \bibfield  {author} {\bibinfo {author} {\bibfnamefont {S.}~\bibnamefont
  {{Weinberg}}},\ }\href {\doibase 10.1103/PhysRevLett.40.223} {\bibfield
  {journal} {\bibinfo  {journal} {Physical Review Letters}\ }\textbf {\bibinfo
  {volume} {40}},\ \bibinfo {pages} {223} (\bibinfo {year} {1978})}\BibitemShut
  {NoStop}%
\bibitem [{\citenamefont {{Wilczek}}(1978)}]{axion3}%
  \BibitemOpen
  \bibfield  {author} {\bibinfo {author} {\bibfnamefont {F.}~\bibnamefont
  {{Wilczek}}},\ }\href {\doibase 10.1103/PhysRevLett.40.279} {\bibfield
  {journal} {\bibinfo  {journal} {Physical Review Letters}\ }\textbf {\bibinfo
  {volume} {40}},\ \bibinfo {pages} {279} (\bibinfo {year} {1978})}\BibitemShut
  {NoStop}%
\bibitem [{\citenamefont {{Kim}}(1979)}]{KSVZ1}%
  \BibitemOpen
  \bibfield  {author} {\bibinfo {author} {\bibfnamefont {J.~E.}\ \bibnamefont
  {{Kim}}},\ }\href {\doibase 10.1103/PhysRevLett.43.103} {\bibfield  {journal}
  {\bibinfo  {journal} {Physical Review Letters}\ }\textbf {\bibinfo {volume}
  {43}},\ \bibinfo {pages} {103} (\bibinfo {year} {1979})}\BibitemShut
  {NoStop}%
\bibitem [{\citenamefont {{Shifman}}\ \emph {et~al.}(1980)\citenamefont
  {{Shifman}}, \citenamefont {{Vainshtein}},\ and\ \citenamefont
  {{Zakharov}}}]{KSVZ2}%
  \BibitemOpen
  \bibfield  {author} {\bibinfo {author} {\bibfnamefont {M.~A.}\ \bibnamefont
  {{Shifman}}}, \bibinfo {author} {\bibfnamefont {A.~I.}\ \bibnamefont
  {{Vainshtein}}}, \ and\ \bibinfo {author} {\bibfnamefont {V.~I.}\
  \bibnamefont {{Zakharov}}},\ }\href {\doibase 10.1016/0550-3213(80)90209-6}
  {\bibfield  {journal} {\bibinfo  {journal} {Nuclear Physics B}\ }\textbf
  {\bibinfo {volume} {166}},\ \bibinfo {pages} {493} (\bibinfo {year}
  {1980})}\BibitemShut {NoStop}%
\bibitem [{\citenamefont {{Dine}}\ \emph {et~al.}(1981)\citenamefont {{Dine}},
  \citenamefont {{Fischler}},\ and\ \citenamefont {{Srednicki}}}]{DFSZ1}%
  \BibitemOpen
  \bibfield  {author} {\bibinfo {author} {\bibfnamefont {M.}~\bibnamefont
  {{Dine}}}, \bibinfo {author} {\bibfnamefont {W.}~\bibnamefont {{Fischler}}},
  \ and\ \bibinfo {author} {\bibfnamefont {M.}~\bibnamefont {{Srednicki}}},\
  }\href {\doibase 10.1016/0370-2693(81)90590-6} {\bibfield  {journal}
  {\bibinfo  {journal} {Physics Letters B}\ }\textbf {\bibinfo {volume}
  {104}},\ \bibinfo {pages} {199} (\bibinfo {year} {1981})}\BibitemShut
  {NoStop}%
\bibitem [{\citenamefont {Zhitnitsky}(1980)}]{DFSZ2}%
  \BibitemOpen
  \bibfield  {author} {\bibinfo {author} {\bibfnamefont {A.~R.}\ \bibnamefont
  {Zhitnitsky}},\ }\href@noop {} {\bibfield  {journal} {\bibinfo  {journal}
  {Sov. J. Nucl. Phys.}\ }\textbf {\bibinfo {volume} {31}},\ \bibinfo {pages}
  {260} (\bibinfo {year} {1980})},\ \bibinfo {note} {[Yad.
  Fiz.31,497(1980)]}\BibitemShut {NoStop}%
\bibitem [{\citenamefont {{Van Bibber}}\ and\ \citenamefont
  {{Rosenberg}}(2006)}]{vanBibber:2006rb}%
  \BibitemOpen
  \bibfield  {author} {\bibinfo {author} {\bibfnamefont {K.}~\bibnamefont {{Van
  Bibber}}}\ and\ \bibinfo {author} {\bibfnamefont {L.~J.}\ \bibnamefont
  {{Rosenberg}}},\ }\href {\doibase 10.1063/1.2349730} {\bibfield  {journal}
  {\bibinfo  {journal} {Physics Today}\ }\textbf {\bibinfo {volume} {59}},\
  \bibinfo {pages} {30} (\bibinfo {year} {2006})}\BibitemShut {NoStop}%
\bibitem [{\citenamefont {{Asztalos}}\ \emph {et~al.}(2006)\citenamefont
  {{Asztalos}}, \citenamefont {{Rosenberg}}, \citenamefont {{van Bibber}},
  \citenamefont {{Sikivie}},\ and\ \citenamefont
  {{Zioutas}}}]{Asztalos:2006kz}%
  \BibitemOpen
  \bibfield  {author} {\bibinfo {author} {\bibfnamefont {S.~J.}\ \bibnamefont
  {{Asztalos}}}, \bibinfo {author} {\bibfnamefont {L.~J.}\ \bibnamefont
  {{Rosenberg}}}, \bibinfo {author} {\bibfnamefont {K.}~\bibnamefont {{van
  Bibber}}}, \bibinfo {author} {\bibfnamefont {P.}~\bibnamefont {{Sikivie}}}, \
  and\ \bibinfo {author} {\bibfnamefont {K.}~\bibnamefont {{Zioutas}}},\ }\href
  {\doibase 10.1146/annurev.nucl.56.080805.140513} {\bibfield  {journal}
  {\bibinfo  {journal} {Annual Review of Nuclear and Particle Science}\
  }\textbf {\bibinfo {volume} {56}},\ \bibinfo {pages} {293} (\bibinfo {year}
  {2006})}\BibitemShut {NoStop}%
\bibitem [{\citenamefont {{Sikivie}}(2008)}]{Sikivie:2008}%
  \BibitemOpen
  \bibfield  {author} {\bibinfo {author} {\bibfnamefont {P.}~\bibnamefont
  {{Sikivie}}},\ }in\ \href@noop {} {\emph {\bibinfo {booktitle} {Axions}}},\
  \bibinfo {series} {Lecture Notes in Physics, Berlin Springer Verlag}, Vol.\
  \bibinfo {volume} {741},\ \bibinfo {editor} {edited by\ \bibinfo {editor}
  {\bibfnamefont {M.}~\bibnamefont {{Kuster}}}, \bibinfo {editor}
  {\bibfnamefont {G.}~\bibnamefont {{Raffelt}}}, \ and\ \bibinfo {editor}
  {\bibfnamefont {B.}~\bibnamefont {{Beltr{\'a}n}}}}\ (\bibinfo {year} {2008})\
  p.~\bibinfo {pages} {19},\ \Eprint {http://arxiv.org/abs/astro-ph/0610440}
  {astro-ph/0610440} \BibitemShut {NoStop}%
\bibitem [{\citenamefont {{Raffelt}}(2008)}]{Raffelt:2006cw}%
  \BibitemOpen
  \bibfield  {author} {\bibinfo {author} {\bibfnamefont {G.~G.}\ \bibnamefont
  {{Raffelt}}},\ }in\ \href@noop {} {\emph {\bibinfo {booktitle} {Axions}}},\
  \bibinfo {series} {Lecture Notes in Physics, Berlin Springer Verlag}, Vol.\
  \bibinfo {volume} {741},\ \bibinfo {editor} {edited by\ \bibinfo {editor}
  {\bibfnamefont {M.}~\bibnamefont {{Kuster}}}, \bibinfo {editor}
  {\bibfnamefont {G.}~\bibnamefont {{Raffelt}}}, \ and\ \bibinfo {editor}
  {\bibfnamefont {B.}~\bibnamefont {{Beltr{\'a}n}}}}\ (\bibinfo {year} {2008})\
  p.~\bibinfo {pages} {51},\ \Eprint {http://arxiv.org/abs/hep-ph/0611350}
  {hep-ph/0611350} \BibitemShut {NoStop}%
\bibitem [{\citenamefont {{Sikivie}}(2010)}]{Sikivie:2009fv}%
  \BibitemOpen
  \bibfield  {author} {\bibinfo {author} {\bibfnamefont {P.}~\bibnamefont
  {{Sikivie}}},\ }\href {\doibase 10.1142/S0217751X10048846} {\bibfield
  {journal} {\bibinfo  {journal} {International Journal of Modern Physics A}\
  }\textbf {\bibinfo {volume} {25}},\ \bibinfo {pages} {554} (\bibinfo {year}
  {2010})},\ \Eprint {http://arxiv.org/abs/0909.0949} {arXiv:0909.0949
  [hep-ph]} \BibitemShut {NoStop}%
\bibitem [{\citenamefont {{Rosenberg}}(2015)}]{Rosenberg:2015kxa}%
  \BibitemOpen
  \bibfield  {author} {\bibinfo {author} {\bibfnamefont {L.~J.}\ \bibnamefont
  {{Rosenberg}}},\ }\href {\doibase 10.1073/pnas.1308788112} {\bibfield
  {journal} {\bibinfo  {journal} {Proceedings of the National Academy of
  Science}\ }\textbf {\bibinfo {volume} {112}},\ \bibinfo {pages} {12278}
  (\bibinfo {year} {2015})}\BibitemShut {NoStop}%
\bibitem [{\citenamefont {{Marsh}}(2016)}]{Marsh:2015xka}%
  \BibitemOpen
  \bibfield  {author} {\bibinfo {author} {\bibfnamefont {D.~J.~E.}\
  \bibnamefont {{Marsh}}},\ }\href {\doibase 10.1016/j.physrep.2016.06.005}
  {\bibfield  {journal} {\bibinfo  {journal} {Physics Reports}\ }\textbf
  {\bibinfo {volume} {643}},\ \bibinfo {pages} {1} (\bibinfo {year} {2016})},\
  \Eprint {http://arxiv.org/abs/1510.07633} {arXiv:1510.07633} \BibitemShut
  {NoStop}%
\bibitem [{\citenamefont {{Graham}}\ \emph {et~al.}(2015)\citenamefont
  {{Graham}}, \citenamefont {{Irastorza}}, \citenamefont {{Lamoreaux}},
  \citenamefont {{Lindner}},\ and\ \citenamefont {{van
  Bibber}}}]{Graham:2015ouw}%
  \BibitemOpen
  \bibfield  {author} {\bibinfo {author} {\bibfnamefont {P.~W.}\ \bibnamefont
  {{Graham}}}, \bibinfo {author} {\bibfnamefont {I.~G.}\ \bibnamefont
  {{Irastorza}}}, \bibinfo {author} {\bibfnamefont {S.~K.}\ \bibnamefont
  {{Lamoreaux}}}, \bibinfo {author} {\bibfnamefont {A.}~\bibnamefont
  {{Lindner}}}, \ and\ \bibinfo {author} {\bibfnamefont {K.~A.}\ \bibnamefont
  {{van Bibber}}},\ }\href {\doibase 10.1146/annurev-nucl-102014-022120}
  {\bibfield  {journal} {\bibinfo  {journal} {Annual Review of Nuclear and
  Particle Science}\ }\textbf {\bibinfo {volume} {65}},\ \bibinfo {pages} {485}
  (\bibinfo {year} {2015})},\ \Eprint {http://arxiv.org/abs/1602.00039}
  {arXiv:1602.00039 [hep-ex]} \BibitemShut {NoStop}%
\bibitem [{\citenamefont {Irastorza}\ and\ \citenamefont
  {Redondo}(2018)}]{Irastorza:2018dyq}%
  \BibitemOpen
  \bibfield  {author} {\bibinfo {author} {\bibfnamefont {I.~G.}\ \bibnamefont
  {Irastorza}}\ and\ \bibinfo {author} {\bibfnamefont {J.}~\bibnamefont
  {Redondo}},\ }\href {\doibase 10.1016/j.ppnp.2018.05.003} {\bibfield
  {journal} {\bibinfo  {journal} {Prog. Part. Nucl. Phys.}\ }\textbf {\bibinfo
  {volume} {102}},\ \bibinfo {pages} {89} (\bibinfo {year} {2018})},\ \Eprint
  {http://arxiv.org/abs/1801.08127} {arXiv:1801.08127 [hep-ph]} \BibitemShut
  {NoStop}%
\bibitem [{\citenamefont {{Liang}}\ and\ \citenamefont
  {{Zhitnitsky}}(2016)}]{Liang:2016tqc}%
  \BibitemOpen
  \bibfield  {author} {\bibinfo {author} {\bibfnamefont {X.}~\bibnamefont
  {{Liang}}}\ and\ \bibinfo {author} {\bibfnamefont {A.}~\bibnamefont
  {{Zhitnitsky}}},\ }\href {\doibase 10.1103/PhysRevD.94.083502} {\bibfield
  {journal} {\bibinfo  {journal} {\prd}\ }\textbf {\bibinfo {volume} {94}},\
  \bibinfo {eid} {083502} (\bibinfo {year} {2016})},\ \Eprint
  {http://arxiv.org/abs/1606.00435} {arXiv:1606.00435 [hep-ph]} \BibitemShut
  {NoStop}%
\bibitem [{\citenamefont {{Ge}}\ \emph {et~al.}(2017)\citenamefont {{Ge}},
  \citenamefont {{Liang}},\ and\ \citenamefont {{Zhitnitsky}}}]{Ge:2017ttc}%
  \BibitemOpen
  \bibfield  {author} {\bibinfo {author} {\bibfnamefont {S.}~\bibnamefont
  {{Ge}}}, \bibinfo {author} {\bibfnamefont {X.}~\bibnamefont {{Liang}}}, \
  and\ \bibinfo {author} {\bibfnamefont {A.}~\bibnamefont {{Zhitnitsky}}},\
  }\href {\doibase 10.1103/PhysRevD.96.063514} {\bibfield  {journal} {\bibinfo
  {journal} {\prd}\ }\textbf {\bibinfo {volume} {96}},\ \bibinfo {eid} {063514}
  (\bibinfo {year} {2017})},\ \Eprint {http://arxiv.org/abs/1702.04354}
  {arXiv:1702.04354 [hep-ph]} \BibitemShut {NoStop}%
\bibitem [{\citenamefont {{Ge}}\ \emph {et~al.}(2018)\citenamefont {{Ge}},
  \citenamefont {{Liang}},\ and\ \citenamefont {{Zhitnitsky}}}]{Ge:2017idw}%
  \BibitemOpen
  \bibfield  {author} {\bibinfo {author} {\bibfnamefont {S.}~\bibnamefont
  {{Ge}}}, \bibinfo {author} {\bibfnamefont {X.}~\bibnamefont {{Liang}}}, \
  and\ \bibinfo {author} {\bibfnamefont {A.}~\bibnamefont {{Zhitnitsky}}},\
  }\href {\doibase 10.1103/PhysRevD.97.043008} {\bibfield  {journal} {\bibinfo
  {journal} {\prd}\ }\textbf {\bibinfo {volume} {97}},\ \bibinfo {eid} {043008}
  (\bibinfo {year} {2018})},\ \Eprint {http://arxiv.org/abs/1711.06271}
  {arXiv:1711.06271 [hep-ph]} \BibitemShut {NoStop}%
\bibitem [{\citenamefont {Ge}\ \emph {et~al.}(2019)\citenamefont {Ge},
  \citenamefont {Lawson},\ and\ \citenamefont {Zhitnitsky}}]{Ge:2019voa}%
  \BibitemOpen
  \bibfield  {author} {\bibinfo {author} {\bibfnamefont {S.}~\bibnamefont
  {Ge}}, \bibinfo {author} {\bibfnamefont {K.}~\bibnamefont {Lawson}}, \ and\
  \bibinfo {author} {\bibfnamefont {A.}~\bibnamefont {Zhitnitsky}},\ }\href
  {\doibase 10.1103/PhysRevD.99.116017} {\bibfield  {journal} {\bibinfo
  {journal} {Phys. Rev. D}\ }\textbf {\bibinfo {volume} {99}},\ \bibinfo
  {pages} {116017} (\bibinfo {year} {2019})},\ \Eprint
  {http://arxiv.org/abs/1903.05090} {arXiv:1903.05090 [hep-ph]} \BibitemShut
  {NoStop}%
\bibitem [{\citenamefont {{Oaknin}}\ and\ \citenamefont
  {{Zhitnitsky}}(2005)}]{Oaknin:2004mn}%
  \BibitemOpen
  \bibfield  {author} {\bibinfo {author} {\bibfnamefont {D.~H.}\ \bibnamefont
  {{Oaknin}}}\ and\ \bibinfo {author} {\bibfnamefont {A.~R.}\ \bibnamefont
  {{Zhitnitsky}}},\ }\href {\doibase 10.1103/PhysRevLett.94.101301} {\bibfield
  {journal} {\bibinfo  {journal} {Physical Review Letters}\ }\textbf {\bibinfo
  {volume} {94}},\ \bibinfo {eid} {101301} (\bibinfo {year} {2005})},\ \Eprint
  {http://arxiv.org/abs/hep-ph/0406146} {hep-ph/0406146} \BibitemShut {NoStop}%
\bibitem [{\citenamefont {{Zhitnitsky}}(2007)}]{Zhitnitsky:2006tu}%
  \BibitemOpen
  \bibfield  {author} {\bibinfo {author} {\bibfnamefont {A.}~\bibnamefont
  {{Zhitnitsky}}},\ }\href {\doibase 10.1103/PhysRevD.76.103518} {\bibfield
  {journal} {\bibinfo  {journal} {\prd}\ }\textbf {\bibinfo {volume} {76}},\
  \bibinfo {eid} {103518} (\bibinfo {year} {2007})},\ \Eprint
  {http://arxiv.org/abs/astro-ph/0607361} {astro-ph/0607361} \BibitemShut
  {NoStop}%
\bibitem [{\citenamefont {{McNeil Forbes}}\ and\ \citenamefont
  {{Zhitnitsky}}(2008)}]{Forbes:2006ba}%
  \BibitemOpen
  \bibfield  {author} {\bibinfo {author} {\bibfnamefont {M.}~\bibnamefont
  {{McNeil Forbes}}}\ and\ \bibinfo {author} {\bibfnamefont {A.~R.}\
  \bibnamefont {{Zhitnitsky}}},\ }\href {\doibase
  10.1088/1475-7516/2008/01/023} {\bibfield  {journal} {\bibinfo  {journal}
  {\jcap}\ }\textbf {\bibinfo {volume} {1}},\ \bibinfo {eid} {023} (\bibinfo
  {year} {2008})},\ \Eprint {http://arxiv.org/abs/astro-ph/0611506}
  {astro-ph/0611506} \BibitemShut {NoStop}%
\bibitem [{\citenamefont {{Lawson}}\ and\ \citenamefont
  {{Zhitnitsky}}(2008)}]{Lawson:2007kp}%
  \BibitemOpen
  \bibfield  {author} {\bibinfo {author} {\bibfnamefont {K.}~\bibnamefont
  {{Lawson}}}\ and\ \bibinfo {author} {\bibfnamefont {A.~R.}\ \bibnamefont
  {{Zhitnitsky}}},\ }\href {\doibase 10.1088/1475-7516/2008/01/022} {\bibfield
  {journal} {\bibinfo  {journal} {\jcap}\ }\textbf {\bibinfo {volume} {1}},\
  \bibinfo {eid} {022} (\bibinfo {year} {2008})},\ \Eprint
  {http://arxiv.org/abs/0704.3064} {arXiv:0704.3064} \BibitemShut {NoStop}%
\bibitem [{\citenamefont {{Forbes}}\ and\ \citenamefont
  {{Zhitnitsky}}(2008)}]{Forbes:2008uf}%
  \BibitemOpen
  \bibfield  {author} {\bibinfo {author} {\bibfnamefont {M.~M.}\ \bibnamefont
  {{Forbes}}}\ and\ \bibinfo {author} {\bibfnamefont {A.~R.}\ \bibnamefont
  {{Zhitnitsky}}},\ }\href {\doibase 10.1103/PhysRevD.78.083505} {\bibfield
  {journal} {\bibinfo  {journal} {\prd}\ }\textbf {\bibinfo {volume} {78}},\
  \bibinfo {eid} {083505} (\bibinfo {year} {2008})},\ \Eprint
  {http://arxiv.org/abs/0802.3830} {arXiv:0802.3830} \BibitemShut {NoStop}%
\bibitem [{\citenamefont {{Forbes}}\ \emph {et~al.}(2010)\citenamefont
  {{Forbes}}, \citenamefont {{Lawson}},\ and\ \citenamefont
  {{Zhitnitsky}}}]{Forbes:2009wg}%
  \BibitemOpen
  \bibfield  {author} {\bibinfo {author} {\bibfnamefont {M.~M.}\ \bibnamefont
  {{Forbes}}}, \bibinfo {author} {\bibfnamefont {K.}~\bibnamefont {{Lawson}}},
  \ and\ \bibinfo {author} {\bibfnamefont {A.~R.}\ \bibnamefont
  {{Zhitnitsky}}},\ }\href {\doibase 10.1103/PhysRevD.82.083510} {\bibfield
  {journal} {\bibinfo  {journal} {\prd}\ }\textbf {\bibinfo {volume} {82}},\
  \bibinfo {eid} {083510} (\bibinfo {year} {2010})},\ \Eprint
  {http://arxiv.org/abs/0910.4541} {arXiv:0910.4541} \BibitemShut {NoStop}%
\bibitem [{\citenamefont {Flambaum}\ and\ \citenamefont
  {Zhitnitsky}(2019)}]{Flambaum:2018ohm}%
  \BibitemOpen
  \bibfield  {author} {\bibinfo {author} {\bibfnamefont {V.~V.}\ \bibnamefont
  {Flambaum}}\ and\ \bibinfo {author} {\bibfnamefont {A.~R.}\ \bibnamefont
  {Zhitnitsky}},\ }\href {\doibase 10.1103/PhysRevD.99.023517} {\bibfield
  {journal} {\bibinfo  {journal} {Phys. Rev. D}\ }\textbf {\bibinfo {volume}
  {99}},\ \bibinfo {pages} {023517} (\bibinfo {year} {2019})},\ \Eprint
  {http://arxiv.org/abs/1811.01965} {arXiv:1811.01965 [hep-ph]} \BibitemShut
  {NoStop}%
\bibitem [{\citenamefont
  {Zhitnitsky}(2020{\natexlab{a}})}]{Zhitnitsky:2019tbh}%
  \BibitemOpen
  \bibfield  {author} {\bibinfo {author} {\bibfnamefont {A.}~\bibnamefont
  {Zhitnitsky}},\ }\href {\doibase 10.1103/PhysRevD.101.083020} {\bibfield
  {journal} {\bibinfo  {journal} {Phys. Rev. D}\ }\textbf {\bibinfo {volume}
  {101}},\ \bibinfo {pages} {083020} (\bibinfo {year} {2020}{\natexlab{a}})},\
  \Eprint {http://arxiv.org/abs/1909.05320} {arXiv:1909.05320 [hep-ph]}
  \BibitemShut {NoStop}%
\bibitem [{\citenamefont {Fraser}\ \emph {et~al.}(2014)\citenamefont {Fraser},
  \citenamefont {Read}, \citenamefont {Sembay}, \citenamefont {Carter},\ and\
  \citenamefont {Schyns}}]{Fraser:2014wja}%
  \BibitemOpen
  \bibfield  {author} {\bibinfo {author} {\bibfnamefont {G.~W.}\ \bibnamefont
  {Fraser}}, \bibinfo {author} {\bibfnamefont {A.~M.}\ \bibnamefont {Read}},
  \bibinfo {author} {\bibfnamefont {S.}~\bibnamefont {Sembay}}, \bibinfo
  {author} {\bibfnamefont {J.~A.}\ \bibnamefont {Carter}}, \ and\ \bibinfo
  {author} {\bibfnamefont {E.}~\bibnamefont {Schyns}},\ }\href {\doibase
  10.1093/mnras/stu1865} {\bibfield  {journal} {\bibinfo  {journal} {Mon. Not.
  Roy. Astron. Soc.}\ }\textbf {\bibinfo {volume} {445}},\ \bibinfo {pages}
  {2146} (\bibinfo {year} {2014})},\ \Eprint {http://arxiv.org/abs/1403.2436}
  {arXiv:1403.2436 [astro-ph.HE]} \BibitemShut {NoStop}%
\bibitem [{\citenamefont {Ge}\ \emph {et~al.}(2020)\citenamefont {Ge},
  \citenamefont {Rachmat}, \citenamefont {Siddiqui}, \citenamefont
  {Van~Waerbeke},\ and\ \citenamefont {Zhitnitsky}}]{Ge:2020cho}%
  \BibitemOpen
  \bibfield  {author} {\bibinfo {author} {\bibfnamefont {S.}~\bibnamefont
  {Ge}}, \bibinfo {author} {\bibfnamefont {H.}~\bibnamefont {Rachmat}},
  \bibinfo {author} {\bibfnamefont {M.~S.~R.}\ \bibnamefont {Siddiqui}},
  \bibinfo {author} {\bibfnamefont {L.}~\bibnamefont {Van~Waerbeke}}, \ and\
  \bibinfo {author} {\bibfnamefont {A.}~\bibnamefont {Zhitnitsky}},\
  }\href@noop {} {\  (\bibinfo {year} {2020})},\ \Eprint
  {http://arxiv.org/abs/2004.00632} {arXiv:2004.00632 [astro-ph.HE]}
  \BibitemShut {NoStop}%
\bibitem [{\citenamefont {Budker}\ \emph {et~al.}(2020)\citenamefont {Budker},
  \citenamefont {Flambaum},\ and\ \citenamefont {Zhitnitsky}}]{Budker:2020mqk}%
  \BibitemOpen
  \bibfield  {author} {\bibinfo {author} {\bibfnamefont {D.}~\bibnamefont
  {Budker}}, \bibinfo {author} {\bibfnamefont {V.~V.}\ \bibnamefont
  {Flambaum}}, \ and\ \bibinfo {author} {\bibfnamefont {A.}~\bibnamefont
  {Zhitnitsky}},\ }\href@noop {} {\  (\bibinfo {year} {2020})},\ \Eprint
  {http://arxiv.org/abs/2003.07363} {arXiv:2003.07363 [hep-ph]} \BibitemShut
  {NoStop}%
\bibitem [{\citenamefont
  {Zhitnitsky}(2020{\natexlab{b}})}]{Zhitnitsky:2020shd}%
  \BibitemOpen
  \bibfield  {author} {\bibinfo {author} {\bibfnamefont {A.}~\bibnamefont
  {Zhitnitsky}},\ }\href@noop {} {\  (\bibinfo {year} {2020}{\natexlab{b}})},\
  \Eprint {http://arxiv.org/abs/2008.04325} {arXiv:2008.04325 [hep-ph]}
  \BibitemShut {NoStop}%
\bibitem [{\citenamefont {{Jacobs}}\ \emph {et~al.}(2015)\citenamefont
  {{Jacobs}}, \citenamefont {{Starkman}},\ and\ \citenamefont
  {{Lynn}}}]{Jacobs:2014yca}%
  \BibitemOpen
  \bibfield  {author} {\bibinfo {author} {\bibfnamefont {D.~M.}\ \bibnamefont
  {{Jacobs}}}, \bibinfo {author} {\bibfnamefont {G.~D.}\ \bibnamefont
  {{Starkman}}}, \ and\ \bibinfo {author} {\bibfnamefont {B.~W.}\ \bibnamefont
  {{Lynn}}},\ }\href {\doibase 10.1093/mnras/stv774} {\bibfield  {journal}
  {\bibinfo  {journal} {\mnras}\ }\textbf {\bibinfo {volume} {450}},\ \bibinfo
  {pages} {3418} (\bibinfo {year} {2015})},\ \Eprint
  {http://arxiv.org/abs/1410.2236} {arXiv:1410.2236} \BibitemShut {NoStop}%
\bibitem [{\citenamefont {Lawson}\ \emph {et~al.}(2019)\citenamefont {Lawson},
  \citenamefont {Liang}, \citenamefont {Mead}, \citenamefont {Siddiqui},
  \citenamefont {Van~Waerbeke},\ and\ \citenamefont
  {Zhitnitsky}}]{Lawson:2019cvy}%
  \BibitemOpen
  \bibfield  {author} {\bibinfo {author} {\bibfnamefont {K.}~\bibnamefont
  {Lawson}}, \bibinfo {author} {\bibfnamefont {X.}~\bibnamefont {Liang}},
  \bibinfo {author} {\bibfnamefont {A.}~\bibnamefont {Mead}}, \bibinfo {author}
  {\bibfnamefont {M.~S.~R.}\ \bibnamefont {Siddiqui}}, \bibinfo {author}
  {\bibfnamefont {L.}~\bibnamefont {Van~Waerbeke}}, \ and\ \bibinfo {author}
  {\bibfnamefont {A.}~\bibnamefont {Zhitnitsky}},\ }\href {\doibase
  10.1103/PhysRevD.100.043531} {\bibfield  {journal} {\bibinfo  {journal}
  {Phys. Rev. D}\ }\textbf {\bibinfo {volume} {100}},\ \bibinfo {pages}
  {043531} (\bibinfo {year} {2019})},\ \Eprint
  {http://arxiv.org/abs/1905.00022} {arXiv:1905.00022 [astro-ph.CO]}
  \BibitemShut {NoStop}%
\bibitem [{\citenamefont {Herrin}\ \emph {et~al.}(2006)\citenamefont {Herrin},
  \citenamefont {Rosenbaum},\ and\ \citenamefont {Teplitz}}]{Herrin:2005kb}%
  \BibitemOpen
  \bibfield  {author} {\bibinfo {author} {\bibfnamefont {E.~T.}\ \bibnamefont
  {Herrin}}, \bibinfo {author} {\bibfnamefont {D.~C.}\ \bibnamefont
  {Rosenbaum}}, \ and\ \bibinfo {author} {\bibfnamefont {V.~L.}\ \bibnamefont
  {Teplitz}},\ }\href {\doibase 10.1103/PhysRevD.73.043511} {\bibfield
  {journal} {\bibinfo  {journal} {Phys. Rev. D}\ }\textbf {\bibinfo {volume}
  {73}},\ \bibinfo {pages} {043511} (\bibinfo {year} {2006})},\ \Eprint
  {http://arxiv.org/abs/astro-ph/0505584} {arXiv:astro-ph/0505584} \BibitemShut
  {NoStop}%
\bibitem [{\citenamefont {Singh~Sidhu}\ \emph {et~al.}(2020)\citenamefont
  {Singh~Sidhu}, \citenamefont {Scherrer},\ and\ \citenamefont
  {Starkman}}]{SinghSidhu:2020cxw}%
  \BibitemOpen
  \bibfield  {author} {\bibinfo {author} {\bibfnamefont {J.}~\bibnamefont
  {Singh~Sidhu}}, \bibinfo {author} {\bibfnamefont {R.~J.}\ \bibnamefont
  {Scherrer}}, \ and\ \bibinfo {author} {\bibfnamefont {G.}~\bibnamefont
  {Starkman}},\ }\href@noop {} {\  (\bibinfo {year} {2020})},\ \Eprint
  {http://arxiv.org/abs/2006.01200} {arXiv:2006.01200 [astro-ph.CO]}
  \BibitemShut {NoStop}%
\bibitem [{\citenamefont {{Parker}}(1988)}]{Parker}%
  \BibitemOpen
  \bibfield  {author} {\bibinfo {author} {\bibfnamefont {E.~N.}\ \bibnamefont
  {{Parker}}},\ }\href {\doibase 10.1086/166485} {\bibfield  {journal}
  {\bibinfo  {journal} {\apj}\ }\textbf {\bibinfo {volume} {330}},\ \bibinfo
  {pages} {474} (\bibinfo {year} {1988})}\BibitemShut {NoStop}%
\bibitem [{\citenamefont {{Krucker}}\ and\ \citenamefont
  {{Benz}}(2000)}]{Benz-2000}%
  \BibitemOpen
  \bibfield  {author} {\bibinfo {author} {\bibfnamefont {S.}~\bibnamefont
  {{Krucker}}}\ and\ \bibinfo {author} {\bibfnamefont {A.~O.}\ \bibnamefont
  {{Benz}}},\ }\href {\doibase 10.1023/A:1005255608792} {\bibfield  {journal}
  {\bibinfo  {journal} {\solphys}\ }\textbf {\bibinfo {volume} {191}},\
  \bibinfo {pages} {341} (\bibinfo {year} {2000})},\ \Eprint
  {http://arxiv.org/abs/astro-ph/9912501} {astro-ph/9912501} \BibitemShut
  {NoStop}%
\bibitem [{\citenamefont {{Benz}}\ and\ \citenamefont
  {{Krucker}}(2001)}]{Benz-2001}%
  \BibitemOpen
  \bibfield  {author} {\bibinfo {author} {\bibfnamefont {A.~O.}\ \bibnamefont
  {{Benz}}}\ and\ \bibinfo {author} {\bibfnamefont {S.}~\bibnamefont
  {{Krucker}}},\ }in\ \href@noop {} {\emph {\bibinfo {booktitle} {Recent
  Insights into the Physics of the Sun and Heliosphere: Highlights from SOHO
  and Other Space Missions}}},\ \bibinfo {series} {IAU Symposium}, Vol.\
  \bibinfo {volume} {203},\ \bibinfo {editor} {edited by\ \bibinfo {editor}
  {\bibfnamefont {P.}~\bibnamefont {{Brekke}}}, \bibinfo {editor}
  {\bibfnamefont {B.}~\bibnamefont {{Fleck}}}, \ and\ \bibinfo {editor}
  {\bibfnamefont {J.~B.}\ \bibnamefont {{Gurman}}}}\ (\bibinfo {year} {2001})\
  p.\ \bibinfo {pages} {471},\ \Eprint {http://arxiv.org/abs/astro-ph/0012106}
  {astro-ph/0012106} \BibitemShut {NoStop}%
\bibitem [{\citenamefont {{Mitra-Kraev}}\ and\ \citenamefont
  {{Benz}}(2001)}]{Kraev-2001}%
  \BibitemOpen
  \bibfield  {author} {\bibinfo {author} {\bibfnamefont {U.}~\bibnamefont
  {{Mitra-Kraev}}}\ and\ \bibinfo {author} {\bibfnamefont {A.~O.}\ \bibnamefont
  {{Benz}}},\ }\href {\doibase 10.1051/0004-6361:20010524} {\bibfield
  {journal} {\bibinfo  {journal} {\aap}\ }\textbf {\bibinfo {volume} {373}},\
  \bibinfo {pages} {318} (\bibinfo {year} {2001})},\ \Eprint
  {http://arxiv.org/abs/astro-ph/0104218} {astro-ph/0104218} \BibitemShut
  {NoStop}%
\bibitem [{\citenamefont {{Benz}}\ and\ \citenamefont
  {{Krucker}}(2002)}]{Benz-2002}%
  \BibitemOpen
  \bibfield  {author} {\bibinfo {author} {\bibfnamefont {A.~O.}\ \bibnamefont
  {{Benz}}}\ and\ \bibinfo {author} {\bibfnamefont {S.}~\bibnamefont
  {{Krucker}}},\ }\href {\doibase 10.1086/338807} {\bibfield  {journal}
  {\bibinfo  {journal} {\apj}\ }\textbf {\bibinfo {volume} {568}},\ \bibinfo
  {pages} {413} (\bibinfo {year} {2002})},\ \Eprint
  {http://arxiv.org/abs/astro-ph/0109027} {astro-ph/0109027} \BibitemShut
  {NoStop}%
\bibitem [{\citenamefont {{Benz}}\ and\ \citenamefont
  {{Grigis}}(2003)}]{Benz-2003}%
  \BibitemOpen
  \bibfield  {author} {\bibinfo {author} {\bibfnamefont {A.~O.}\ \bibnamefont
  {{Benz}}}\ and\ \bibinfo {author} {\bibfnamefont {P.~C.}\ \bibnamefont
  {{Grigis}}},\ }\href {\doibase 10.1016/S0273-1177(03)00306-5} {\bibfield
  {journal} {\bibinfo  {journal} {Advances in Space Research}\ }\textbf
  {\bibinfo {volume} {32}},\ \bibinfo {pages} {1035} (\bibinfo {year}
  {2003})},\ \Eprint {http://arxiv.org/abs/astro-ph/0308323} {astro-ph/0308323}
  \BibitemShut {NoStop}%
\bibitem [{\citenamefont {{Pauluhn}}\ and\ \citenamefont
  {{Solanki}}(2007)}]{Pauluhn:2006ut}%
  \BibitemOpen
  \bibfield  {author} {\bibinfo {author} {\bibfnamefont {A.}~\bibnamefont
  {{Pauluhn}}}\ and\ \bibinfo {author} {\bibfnamefont {S.~K.}\ \bibnamefont
  {{Solanki}}},\ }\href {\doibase 10.1051/0004-6361:20065152} {\bibfield
  {journal} {\bibinfo  {journal} {\aap}\ }\textbf {\bibinfo {volume} {462}},\
  \bibinfo {pages} {311} (\bibinfo {year} {2007})},\ \Eprint
  {http://arxiv.org/abs/astro-ph/0612585} {astro-ph/0612585} \BibitemShut
  {NoStop}%
\bibitem [{\citenamefont {{Hannah}}\ \emph {et~al.}(2008)\citenamefont
  {{Hannah}}, \citenamefont {{Christe}}, \citenamefont {{Krucker}},
  \citenamefont {{Hurford}}, \citenamefont {{Hudson}},\ and\ \citenamefont
  {{Lin}}}]{Hannah:2007kw}%
  \BibitemOpen
  \bibfield  {author} {\bibinfo {author} {\bibfnamefont {I.~G.}\ \bibnamefont
  {{Hannah}}}, \bibinfo {author} {\bibfnamefont {S.}~\bibnamefont {{Christe}}},
  \bibinfo {author} {\bibfnamefont {S.}~\bibnamefont {{Krucker}}}, \bibinfo
  {author} {\bibfnamefont {G.~J.}\ \bibnamefont {{Hurford}}}, \bibinfo {author}
  {\bibfnamefont {H.~S.}\ \bibnamefont {{Hudson}}}, \ and\ \bibinfo {author}
  {\bibfnamefont {R.~P.}\ \bibnamefont {{Lin}}},\ }\href {\doibase
  10.1086/529012} {\bibfield  {journal} {\bibinfo  {journal} {\apj}\ }\textbf
  {\bibinfo {volume} {677}},\ \bibinfo {pages} {704} (\bibinfo {year}
  {2008})},\ \Eprint {http://arxiv.org/abs/0712.2544} {arXiv:0712.2544}
  \BibitemShut {NoStop}%
\bibitem [{\citenamefont {{Bingert}}\ and\ \citenamefont
  {{Peter}}(2013)}]{Bingert:2012se}%
  \BibitemOpen
  \bibfield  {author} {\bibinfo {author} {\bibfnamefont {S.}~\bibnamefont
  {{Bingert}}}\ and\ \bibinfo {author} {\bibfnamefont {H.}~\bibnamefont
  {{Peter}}},\ }\href {\doibase 10.1051/0004-6361/201220469} {\bibfield
  {journal} {\bibinfo  {journal} {\aap}\ }\textbf {\bibinfo {volume} {550}},\
  \bibinfo {eid} {A30} (\bibinfo {year} {2013})},\ \Eprint
  {http://arxiv.org/abs/1211.6417} {arXiv:1211.6417 [astro-ph.SR]} \BibitemShut
  {NoStop}%
\bibitem [{\citenamefont {{Terzo}}\ \emph {et~al.}(2011)\citenamefont
  {{Terzo}}, \citenamefont {{Reale}}, \citenamefont {{Miceli}}, \citenamefont
  {{Klimchuk}}, \citenamefont {{Kano}},\ and\ \citenamefont
  {{Tsuneta}}}]{terzo-2011}%
  \BibitemOpen
  \bibfield  {author} {\bibinfo {author} {\bibfnamefont {S.}~\bibnamefont
  {{Terzo}}}, \bibinfo {author} {\bibfnamefont {F.}~\bibnamefont {{Reale}}},
  \bibinfo {author} {\bibfnamefont {M.}~\bibnamefont {{Miceli}}}, \bibinfo
  {author} {\bibfnamefont {J.~A.}\ \bibnamefont {{Klimchuk}}}, \bibinfo
  {author} {\bibfnamefont {R.}~\bibnamefont {{Kano}}}, \ and\ \bibinfo {author}
  {\bibfnamefont {S.}~\bibnamefont {{Tsuneta}}},\ }\href {\doibase
  10.1088/0004-637X/736/2/111} {\bibfield  {journal} {\bibinfo  {journal}
  {\apj}\ }\textbf {\bibinfo {volume} {736}},\ \bibinfo {eid} {111} (\bibinfo
  {year} {2011})},\ \Eprint {http://arxiv.org/abs/1105.2506} {arXiv:1105.2506
  [astro-ph.SR]} \BibitemShut {NoStop}%
\bibitem [{\citenamefont {{Bradshaw}}\ \emph {et~al.}(2012)\citenamefont
  {{Bradshaw}}, \citenamefont {{Klimchuk}},\ and\ \citenamefont
  {{Reep}}}]{Bradshaw-2012}%
  \BibitemOpen
  \bibfield  {author} {\bibinfo {author} {\bibfnamefont {S.~J.}\ \bibnamefont
  {{Bradshaw}}}, \bibinfo {author} {\bibfnamefont {J.~A.}\ \bibnamefont
  {{Klimchuk}}}, \ and\ \bibinfo {author} {\bibfnamefont {J.~W.}\ \bibnamefont
  {{Reep}}},\ }\href {\doibase 10.1088/0004-637X/758/1/53} {\bibfield
  {journal} {\bibinfo  {journal} {\apj}\ }\textbf {\bibinfo {volume} {758}},\
  \bibinfo {eid} {53} (\bibinfo {year} {2012})},\ \Eprint
  {http://arxiv.org/abs/1209.0737} {arXiv:1209.0737 [astro-ph.SR]} \BibitemShut
  {NoStop}%
\bibitem [{\citenamefont {{Jess}}\ \emph {et~al.}(2014)\citenamefont {{Jess}},
  \citenamefont {{Mathioudakis}},\ and\ \citenamefont {{Keys}}}]{Jess-2014}%
  \BibitemOpen
  \bibfield  {author} {\bibinfo {author} {\bibfnamefont {D.~B.}\ \bibnamefont
  {{Jess}}}, \bibinfo {author} {\bibfnamefont {M.}~\bibnamefont
  {{Mathioudakis}}}, \ and\ \bibinfo {author} {\bibfnamefont {P.~H.}\
  \bibnamefont {{Keys}}},\ }\href {\doibase 10.1088/0004-637X/795/2/172}
  {\bibfield  {journal} {\bibinfo  {journal} {\apj}\ }\textbf {\bibinfo
  {volume} {795}},\ \bibinfo {eid} {172} (\bibinfo {year} {2014})},\ \Eprint
  {http://arxiv.org/abs/1409.6726} {arXiv:1409.6726 [astro-ph.SR]} \BibitemShut
  {NoStop}%
\bibitem [{\citenamefont {{Kirichenko}}\ and\ \citenamefont
  {{Bogachev}}(2017)}]{Kirichenko-2017}%
  \BibitemOpen
  \bibfield  {author} {\bibinfo {author} {\bibfnamefont {A.~S.}\ \bibnamefont
  {{Kirichenko}}}\ and\ \bibinfo {author} {\bibfnamefont {S.~A.}\ \bibnamefont
  {{Bogachev}}},\ }\href {\doibase 10.3847/1538-4357/aa6c2b} {\bibfield
  {journal} {\bibinfo  {journal} {\apj}\ }\textbf {\bibinfo {volume} {840}},\
  \bibinfo {eid} {45} (\bibinfo {year} {2017})},\ \Eprint
  {http://arxiv.org/abs/1706.05852} {arXiv:1706.05852 [astro-ph.SR]}
  \BibitemShut {NoStop}%
\bibitem [{\citenamefont {{Mac Cormack}}\ \emph {et~al.}(2017)\citenamefont
  {{Mac Cormack}}, \citenamefont {{V{\'a}squez}}, \citenamefont {{L{\'o}pez
  Fuentes}}, \citenamefont {{Nuevo}}, \citenamefont {{Landi}},\ and\
  \citenamefont {{Frazin}}}]{Cormack-2017}%
  \BibitemOpen
  \bibfield  {author} {\bibinfo {author} {\bibfnamefont {C.}~\bibnamefont {{Mac
  Cormack}}}, \bibinfo {author} {\bibfnamefont {A.~M.}\ \bibnamefont
  {{V{\'a}squez}}}, \bibinfo {author} {\bibfnamefont {M.}~\bibnamefont
  {{L{\'o}pez Fuentes}}}, \bibinfo {author} {\bibfnamefont {F.~A.}\
  \bibnamefont {{Nuevo}}}, \bibinfo {author} {\bibfnamefont {E.}~\bibnamefont
  {{Landi}}}, \ and\ \bibinfo {author} {\bibfnamefont {R.~A.}\ \bibnamefont
  {{Frazin}}},\ }\href {\doibase 10.3847/1538-4357/aa76e9} {\bibfield
  {journal} {\bibinfo  {journal} {\apj}\ }\textbf {\bibinfo {volume} {843}},\
  \bibinfo {eid} {70} (\bibinfo {year} {2017})},\ \Eprint
  {http://arxiv.org/abs/1706.00365} {arXiv:1706.00365 [astro-ph.SR]}
  \BibitemShut {NoStop}%
\bibitem [{\citenamefont {{Bertolucci}}\ \emph {et~al.}(2017)\citenamefont
  {{Bertolucci}}, \citenamefont {{Zioutas}}, \citenamefont {{Hofmann}},\ and\
  \citenamefont {{Maroudas}}}]{Bertolucci-2017}%
  \BibitemOpen
  \bibfield  {author} {\bibinfo {author} {\bibfnamefont {S.}~\bibnamefont
  {{Bertolucci}}}, \bibinfo {author} {\bibfnamefont {K.}~\bibnamefont
  {{Zioutas}}}, \bibinfo {author} {\bibfnamefont {S.}~\bibnamefont
  {{Hofmann}}}, \ and\ \bibinfo {author} {\bibfnamefont {M.}~\bibnamefont
  {{Maroudas}}},\ }\href {\doibase 10.1016/j.dark.2017.06.001} {\bibfield
  {journal} {\bibinfo  {journal} {Physics of the Dark Universe}\ }\textbf
  {\bibinfo {volume} {17}},\ \bibinfo {pages} {13} (\bibinfo {year} {2017})},\
  \Eprint {http://arxiv.org/abs/1602.03666} {arXiv:1602.03666 [astro-ph.SR]}
  \BibitemShut {NoStop}%
\bibitem [{\citenamefont {{Bingert, S.}}\ and\ \citenamefont {{Peter,
  H.}}(2013)}]{Bingert:2013}%
  \BibitemOpen
  \bibfield  {author} {\bibinfo {author} {\bibnamefont {{Bingert, S.}}}\ and\
  \bibinfo {author} {\bibnamefont {{Peter, H.}}},\ }\href {\doibase
  10.1051/0004-6361/201220469} {\bibfield  {journal} {\bibinfo  {journal}
  {A\&A}\ }\textbf {\bibinfo {volume} {550}},\ \bibinfo {pages} {A30} (\bibinfo
  {year} {2013})}\BibitemShut {NoStop}%
\bibitem [{\citenamefont {Thejappa}(1991)}]{Thejappa-1991}%
  \BibitemOpen
  \bibfield  {author} {\bibinfo {author} {\bibfnamefont {G.}~\bibnamefont
  {Thejappa}},\ }\href {\doibase 10.1007/BF00159137} {\bibfield  {journal}
  {\bibinfo  {journal} {Solar Physics}\ }\textbf {\bibinfo {volume} {132}},\
  \bibinfo {pages} {173} (\bibinfo {year} {1991})}\BibitemShut {NoStop}%
\bibitem [{\citenamefont {Zhitnitsky}(2018)}]{Zhitnitsky:2018mav}%
  \BibitemOpen
  \bibfield  {author} {\bibinfo {author} {\bibfnamefont {A.}~\bibnamefont
  {Zhitnitsky}},\ }\href {\doibase 10.1016/j.dark.2018.08.001} {\bibfield
  {journal} {\bibinfo  {journal} {Phys. Dark Univ.}\ }\textbf {\bibinfo
  {volume} {22}},\ \bibinfo {pages} {1} (\bibinfo {year} {2018})},\ \Eprint
  {http://arxiv.org/abs/1801.01509} {arXiv:1801.01509 [astro-ph.SR]}
  \BibitemShut {NoStop}%
\bibitem [{\citenamefont {{Avrett}}\ and\ \citenamefont
  {{Loeser}}(2008)}]{2008ApJS..175..229A}%
  \BibitemOpen
  \bibfield  {author} {\bibinfo {author} {\bibfnamefont {E.~H.}\ \bibnamefont
  {{Avrett}}}\ and\ \bibinfo {author} {\bibfnamefont {R.}~\bibnamefont
  {{Loeser}}},\ }\href {\doibase 10.1086/523671} {\bibfield  {journal}
  {\bibinfo  {journal} {\apjs}\ }\textbf {\bibinfo {volume} {175}},\ \bibinfo
  {pages} {229} (\bibinfo {year} {2008})}\BibitemShut {NoStop}%
\bibitem [{\citenamefont {Sharma}\ \emph {et~al.}(2018)\citenamefont {Sharma},
  \citenamefont {Oberoi},\ and\ \citenamefont {Arjunwadkar}}]{Sharma_2018}%
  \BibitemOpen
  \bibfield  {author} {\bibinfo {author} {\bibfnamefont {R.}~\bibnamefont
  {Sharma}}, \bibinfo {author} {\bibfnamefont {D.}~\bibnamefont {Oberoi}}, \
  and\ \bibinfo {author} {\bibfnamefont {M.}~\bibnamefont {Arjunwadkar}},\
  }\href {\doibase 10.3847/1538-4357/aa9d96} {\bibfield  {journal} {\bibinfo
  {journal} {The Astrophysical Journal}\ }\textbf {\bibinfo {volume} {852}},\
  \bibinfo {pages} {69} (\bibinfo {year} {2018})}\BibitemShut {NoStop}%
\bibitem [{\citenamefont {Oberoi}\ \emph {et~al.}(2017)\citenamefont {Oberoi},
  \citenamefont {Sharma},\ and\ \citenamefont {Rogers}}]{oberoi_2017}%
  \BibitemOpen
  \bibfield  {author} {\bibinfo {author} {\bibfnamefont {D.}~\bibnamefont
  {Oberoi}}, \bibinfo {author} {\bibfnamefont {R.}~\bibnamefont {Sharma}}, \
  and\ \bibinfo {author} {\bibfnamefont {A.~E.~E.}\ \bibnamefont {Rogers}},\
  }\href {http://dx.doi.org/10.1007/s11207-017-1096-1} {\bibfield  {journal}
  {\bibinfo  {journal} {Solar Physics}\ }\textbf {\bibinfo {volume} {292}},\
  \bibinfo {pages} {1} (\bibinfo {year} {2017})}\BibitemShut {NoStop}%
\bibitem [{\citenamefont {Wheatland}\ \emph {et~al.}(1998)\citenamefont
  {Wheatland}, \citenamefont {Sturrock},\ and\ \citenamefont
  {McTiernan}}]{Wheatland_1998}%
  \BibitemOpen
  \bibfield  {author} {\bibinfo {author} {\bibfnamefont {M.~S.}\ \bibnamefont
  {Wheatland}}, \bibinfo {author} {\bibfnamefont {P.~A.}\ \bibnamefont
  {Sturrock}}, \ and\ \bibinfo {author} {\bibfnamefont {J.~M.}\ \bibnamefont
  {McTiernan}},\ }\href {\doibase 10.1086/306492} {\bibfield  {journal}
  {\bibinfo  {journal} {The Astrophysical Journal}\ }\textbf {\bibinfo {volume}
  {509}},\ \bibinfo {pages} {448} (\bibinfo {year} {1998})}\BibitemShut
  {NoStop}%
\bibitem [{\citenamefont {Aschwanden}\ and\ \citenamefont
  {McTiernan}(2010)}]{Aschwanden_2010}%
  \BibitemOpen
  \bibfield  {author} {\bibinfo {author} {\bibfnamefont {M.~J.}\ \bibnamefont
  {Aschwanden}}\ and\ \bibinfo {author} {\bibfnamefont {J.~M.}\ \bibnamefont
  {McTiernan}},\ }\href {\doibase 10.1088/0004-637x/717/2/683} {\bibfield
  {journal} {\bibinfo  {journal} {The Astrophysical Journal}\ }\textbf
  {\bibinfo {volume} {717}},\ \bibinfo {pages} {683} (\bibinfo {year}
  {2010})}\BibitemShut {NoStop}%
\bibitem [{\citenamefont {Li}\ \emph {et~al.}(2014)\citenamefont {Li},
  \citenamefont {Zhong}, \citenamefont {Wang}, \citenamefont {Su},\ and\
  \citenamefont {Fang}}]{Li_2014}%
  \BibitemOpen
  \bibfield  {author} {\bibinfo {author} {\bibfnamefont {C.}~\bibnamefont
  {Li}}, \bibinfo {author} {\bibfnamefont {S.~J.}\ \bibnamefont {Zhong}},
  \bibinfo {author} {\bibfnamefont {L.}~\bibnamefont {Wang}}, \bibinfo {author}
  {\bibfnamefont {W.}~\bibnamefont {Su}}, \ and\ \bibinfo {author}
  {\bibfnamefont {C.}~\bibnamefont {Fang}},\ }\href {\doibase
  10.1088/2041-8205/792/2/l26} {\bibfield  {journal} {\bibinfo  {journal} {The
  Astrophysical Journal}\ }\textbf {\bibinfo {volume} {792}},\ \bibinfo {pages}
  {L26} (\bibinfo {year} {2014})}\BibitemShut {NoStop}%
\end{thebibliography}%

\end{document}